\documentclass[a4paper,fleqn,usenatbib]{mnras}

\usepackage{newtxtext,newtxmath}
\usepackage[T1]{fontenc}
\usepackage{ae,aecompl}

\usepackage{graphicx}	% Including figure files
\usepackage{amsmath}	% Advanced maths commands
\usepackage{amssymb}	% Extra maths symbols

\usepackage{multirow}
\usepackage{xspace}

\def\aap{A\&A}                % Astronomy and Astrophysics
\def\aapr{A\&A~Rev.}          % Astronomy and Astrophysics Reviews
\def\aj{AJ}                   % Astronomical Journal
\def\apj{ApJ}                 % Astrophysical Journal
\def\apjl{ApJ}                % Astrophysical Journal, Letters
\def\apjs{ApJS}               % Astrophysical Journal, Supplement
\def\araa{ARA\&A}             % Annual Review of Astronomy and Astrophysics
\def\mnras{MNRAS}             % Monthly Notices of the RAS
\def\nat{Nature}              % Nature
\def\pasj{PASJ}               % Publ. of the Astronomical Society of Japan
\def\pasp{PASP}               % Publ. of the Astron. Society of the Pacific
\def\procspie{Proc.~SPIE}     % Proceedings of the SPIE
\def\rosat{{\it ROSAT}\xspace}
\def\asca{{\it ASCA}\xspace}
\def\xmm{{\it XMM-Newton}\xspace}

\def\hst{{\it HST}\xspace}
\def\hubble{{\it HST}\xspace}
\def\chandra{{\it Chandra}\xspace}
\def\spitzer{{\it Spitzer}\xspace}
\def\bepposax{{\it BeppoSAX}\xspace}

\def\emerlin{e-MERLIN\xspace}

\def\ergsec{{\rm erg~s$^{-1}$}\xspace}

\def\chisqdof{$\chi^{2}$/DoF\xspace}

\def\redchisq{$\chi^{2}_{\rm red}$\xspace}

\def\deg{$^{\circ}$\xspace}
\def\H0{{\rm ~km~s^{-1}~Mpc^{-1}}}

\def\la{\mathrel{\hbox{\rlap{\hbox{\lower4pt\hbox{$\sim$}}}{\raise2pt\hbox{$<$}}}}}
\def\ga{\mathrel{\hbox{\rlap{\hbox{\lower4pt\hbox{$\sim$}}}{\raise2pt\hbox{$>$}}}}}

\def\d25{$D_{25}$}
\def\nh{$N_{\rm H}$\xspace}

\def\lx{$L_{\rm X}$}

\def\mbh{$M_{\rm BH}$\xspace}

\def\lbolledd{$L_{\rm bol}/L_{\rm Edd}$\xspace}
\def\lbol{$L_{\rm bol}$\xspace}
\def\lledd{$L/L_{\rm Edd}$\xspace}

\def\ledd{$L_{\rm Edd}$\xspace}

\def\othree{[\ion{O}{iii}]\xspace}
\def\hbeta{H$\beta$\xspace}
\def\ntwo{[\ion{N}{ii}]\xspace}
\def\halpha{H$\alpha$\xspace}
\def\stwo{[\ion{S}{ii}]\xspace}
\def\oone{[\ion{O}{i}]\xspace}

\def\msol{$M_{\odot}$\xspace}

\def\lbol{$L_{\rm bol}$\xspace}

\def\njet{N$_{\rm j}$\xspace}
\def\te{T$_{\rm e}$\xspace}
\def\zacc{$z_{\rm acc}$\xspace}
\def\rin{r$_{\rm in}$\xspace}
\def\rout{r$_{\rm out}$\xspace}
\def\tin{T$_{\rm in}$\xspace}

\def\escat{$\epsilon_{\rm scat}$\xspace}
\def\zmax{$z_{\rm max}$\xspace}
\def\hratio{$h_{\rm 0}/r_{\rm 0}$\xspace}
\def\rg{$r_{\rm g}$\xspace}
\def\rzero{r$_{0}$\xspace}

\def\syn{{\sc syn}\xspace}
\def\syncom{{\sc syncom}\xspace}
\def\com{{\sc com}\xspace}

\def\grs{GRS\,1915+105\xspace}
\def\gx{GX\,339--4\xspace}
\def\ngc{NGC\,4736\xspace}
\def\4051{NGC\,4051\xspace}
\def\m94{M94*\xspace}
\def\a0620{A0620-00\xspace}
\def\xtej{XTE\,J1118+480\xspace}
\def\groj{GRO\,J1655--40\xspace}
\def\sgra{Sgr\,A*\xspace}

\newcommand{\mc}{\multicolumn}
\newcommand{\mr}{\multirow}

\title[A jet model for \m94]
{A jet-dominated model for a broad-band spectral energy distribution of the nearby low-luminosity active galactic nucleus in M94}

\begin{document}

\defcitealias{MarkoffNowakWilms2005}{MNW05}

\author[P.C.N. van Oers et al.]{Pieter van Oers$^{1}$, Sera Markoff$^1$, Phil Uttley$^1$, Ian McHardy$^2$, \and Tessel van der Laan$^3$, Jennifer Donovan Meyer$^4$, Riley Connors$^1$\\
$^1$Anton Pannekoek Institute for Astronomy, University of Amsterdam, Science Park 904, 1098 XH, Amsterdam, The Netherlands\\
$^2$School of Physics and Astronomy, University of Southampton, Southampton, S017 1BJ, UK\\
$^3$Institut de Radioastronomie Millimétrique, 300 rue de la Piscine, 38406 St. Martin d'Hères, Grenoble, France\\
$^4$National Radio Astronomy Observatory, Charlottesville, VA 22901, USA\\
}

%\pagerange{\pageref{firstpage}--\pageref{lastpage}} \pubyear{2008}

\maketitle

\label{firstpage}

\begin{abstract}

We have compiled a new multi-wavelength spectral energy distribution (SED) for the closest obscured low ionisation emission line region (LINER) active galactic nucleus (AGN), \ngc, also known as M94. The SED comprises mainly high resolution (mostly sub-arcsecond, or, at the distance to M94, $\lesssim23$ pc from the nucleus) observations from the literature, archival data, as well as previously unpublished sub-millimetre data from the Plateau de Bure Interferometer (PdBI) and the Combined Array for Research in Millimeter-wave Astronomy (CARMA), in conjunction with new electronic Multi Element Radio Interferometric Network (\emerlin) L-band (1.5 GHz) observations. Thanks to the \emerlin resolution and sensitivity we resolve for the first time a double structure composed of two radio sources separated by $\sim1$\arcsec, previously observed only at higher frequency. We explore this dataset, that further includes non-simultaneous data from the Very Large Array (VLA), the Gemini telescope, the Hubble Space Telescope (\hst) and the \chandra X-ray observatory, in terms of an outflow-dominated model. We compare our results with previous trends found for other AGN using the same model (\4051, M81*, M87 and \sgra), as well as hard and quiescent state X-ray binaries. We find that the nuclear broadband spectrum of M94 is consistent with a relativistic outflow of low inclination. The findings in this work add to the growing body of evidence that the physics of weakly accreting black holes scales with mass in a rather straightforward fashion.

\end{abstract}


%%%%%%%%%%%%%%%%%%%%%%%%%%%%%%%%%%%%%%%%%%%%
%%%%%%%%%%%%%%%%%%%%%%%%%%%%%%%%%%%%%%%%%%%%%%%%%
%INTRO%%%%%%%%%%%%%%%%%%%%%%%%%%%%%%%%%%%%%%
%%%%%%%%%%%%%%%%%%%%%%%%%%%%%%%%%%%%%%%%%%%%%%%%
\section{Introduction}
\label{sec:nintro}

At 4.8$\pm0.8$ Mpc, M94 is the closest known galaxy that has a low-luminosity active galactic nucleus (LLAGN) classified as an \emph{obscured} low ionisation emission line region LINER (L2, \citealt{Ho+1997,Roberts+2001}) and is the least luminous member of the LINER class found thus far (\lbol$\sim2.5\times10^{40}$ \ergsec; \citealt{Constantin+2012}). Therefore the M94 nucleus represents a unique possibility to probe the extremely low-accretion rate physics thought to govern these types of sources. In this work we aim to do a multi-wavelength study of this source and hence we will first briefly describe this source in different wavebands.

M94 has an optical radius $R_{25}$ of $\sim 5.6\arcmin$ or $\sim7$ kpc \citep{Trujillo+2009} and was identified as a member of Hubble class (R)SAB(rs)ab \citep{Hubble1926,Buta+2007}. i.e. an early type spiral galaxy with star-forming rings. The central surface brightness profile in infrared reveals a double S\'{e}rsic structure \citep[][these sources are on average less radio-loud than so-called ``core" galaxies]{RichingsUttleyKording2011}. \citet{Eracleous+2002} observed a complex X-ray structure, harbouring a myriad of discrete X-ray sources. 

\citet{Maoz+1995,Maoz+2005} speculate that the galaxy hosts a double AGN in the process of merging, which may have triggered a starburst episode and could explain the observed morphological and kinematic features. On the other hand, there has been much debate about whether even a single AGN is lurking in M94, as the LINER could just as easily be ionised by e.g. the UV radiation from hot, young stars, collisional shocks or a population of stellar mass black holes (BHs; e.g. \citealt{Roberts+1999,Pellegrini+2002} and references there-in). However evidence that M94 harbours an AGN has been accumulating: The second brightest X-ray source (as seen by \chandra) has counterparts in the UV \citep{Maoz+2005} and radio \citep{KoerdingColbertFalcke2005,NagarFalckeWilson2002,NagarFalckeWilson2005} and hence seems to correspond to a supermassive BH (SMBH). Further evidence for the AGN nature of this source includes e.g. its variability in the UV \citep{Maoz+2005}, and more recently the analysis of archival \hubble Space Telescope Imaging Spectrograph (STIS) data by \citet{Constantin+2012}, who discovered a broad H$\alpha$ emission line in the nuclear optical spectrum; generally considered \emph{the} ``smoking gun" for AGN emission. 

Perhaps the best evidence for what is the dominant source of ionisation in M94 comes from detailed optical spectroscopic analysis of the emission lines originating from the narrow line region (NLR). Emission line diagnostic diagrams (generally referred to as BPT diagrams, after \citealt{BaldwinPhillipsTerlevich1981}) make use of the ratios of optical lines (\ntwo/\halpha, \othree/\hbeta) to separate galaxies dominated by star formation (SF) from those dominated by AGN emission \citep[also see][]{VeilleuxOsterbrock1987,Kauffmann+2003,Kewley+2006,Schawinski+2007}. When the \stwo and \oone lines are also detected, the BPT diagrams allow us to accurately classify the cores of galaxies that are not dominated by stellar processes into two main types of AGN: LINERs and Seyferts. The former are generally of lower bolometric luminosity (\lbol) and higher radio loudness, while the latter are historically associated with higher \lbol, radio quiet sources (but see e.g. \citealt{UlvestadWilson1989,HoPeng2001,Maitra+2011} and references therein). Figure 2 in \citet{Constantin+2012} summarises the currently available M94 nuclear optical spectroscopy and shows it is consistent with a LINER \citep[also see][]{MoustakasKennicutt2006}. 

It is a topic of ongoing investigation whether the two different types of AGN mentioned above may actually correspond or ``map" to certain states found in their low-mass relatives, the BHs in X-ray binaries (i.e. BH binaries; BHBs) \citep{MerloniHeinzDiMatteo2003,Falcke+2004,KoerdingJesterFender2006,McHardy+2006,Plotkin+2012}. We observe BHBs in predominantly two spectral states, named after their appearance in X-rays (see, e.g. \citealt{RemillardMcClintock2006} or \citealt{Belloni2010} for reviews on BHB spectral states). The soft state X-ray spectrum is dominated by a soft, thermal component, usually associated with a geometrically thin, optically thick accretion disk (\citealt{NovikovThorne1973,ShakuraSunyaev1973}). The Hard State (HS) is dominated by an optically thin, hard X-ray component, that is usually thought to be due to inverse Comptonisation of a thermal input \citep{ThornePrice1975,SunyaevTruemper1979}. In addition to this power-law-like emission, the HS also displays high RMS noise and the presence of compact, steady jets. Hence on the surface it is tempting to associate the LINER class with the HS and the Seyfert class with the soft state of BHBs however see Sec. \ref{sec:disccomp}), but to see how far this ostensible comparison would hold or fail, more quantitative comparisons are necessary.

While BHBs can typically change state on humanly observable timescales of days to weeks, their supermassive counterparts are much slower to go through their evolution because of the length scales involved, with typical timescales expected for state changes easily exceeding $10^4$ years (e.g. \citealt{Keel+2012,Keel+2015,Schawinski+2015} and references therein). Hence it is nearly impossible to directly observe state changes in AGN. Interestingly, however, AGN that show significant variation in the intrinsic luminosity have been observed \citep[see e.g.][and references there-in]{LaMassa+2015,Koay+2016}. In addition, for the first time a quasar was found to ``switch off", possibly due to a transition into a radiatively inefficient state \citep{Schawinski+2010}. More indirect evidence that AGN can change state comes from X-ray images of cavities in clusters of galaxies, showing activity cycles of $\sim10^7$--$10^8$ years \citep[e.g.][]{Birzan+2004}. Clearly such findings show the merits of studying and comparing the evolution of BHs on both ends of the mass-scale, specifically for the predictive nature of low mass BHB evolution when used to infer AGN evolution.

Low accretion rate BHBs are generally thought to harbour a radiatively inefficient accretion flow (RIAF). In such flows a thermal-dominated ``cloud" of electrons comprise the inner flow. This cloud can be seen as the inner accretion disk that evaporated into a ``corona" \citep{Esin+1997,Gilfanov2010}, or can be interpreted as the base of a jet \citep[][hereafter \citetalias{MarkoffNowakWilms2005}]{MarkoffNowakWilms2005}, i.e. a magnetised, outflowing corona (e.g. \citealt{Beloborodov1999,Malzac+2001,MerloniFabian2002}). In this work we explore the concept of mass-scaling in BH systems, investigating a newly compiled broadband Spectral Energy Distribution (SED) of the M94 nucleus in the context of an outflow-dominated relativistic jet paradigm \citepalias[][see Section \ref{sec:model} for further references]{MarkoffNowakWilms2005}. The SED consists of some of the first data acquired with the recently upgraded \emerlin, supplemented by previously unpublished Plateau de Bure Interferometer (PdBI) and Combined Array for Research in Millimeter-wave Astronomy (CARMA) continuum data, as well as archival data and data from the literature. The \citetalias{MarkoffNowakWilms2005} framework is, to date, the only model that has been applied successfully to a variety of BH \emph{broadband} SEDs on both ends of the mass scale, as well as across the luminosity scale (in terms of the Eddington luminosity, \lbolledd, where \ledd = $1.26\times10^{38}$ [\mbh/M$_{\odot}$] \ergsec, with \mbh the BH mass). Specifically the observations investigated with \citetalias{MarkoffNowakWilms2005} range from a multitude of HS and quiescent state BHBs \citep[\citetalias{MarkoffNowakWilms2005};][]{Gallo+2007,Migliari+2007,MaitraMarkoffBrocksopp2009,Plotkin+2015}, to the \grs specific (HS equivalent) $\chi$ state \citep{vanOers+2010} to several SMBHs \citep[][in prep.]{MarkoffBowerFalcke2007,Markoff+2008,Maitra+2009b,Maitra+2011,Prieto+2016,Connors+2016}. Consistent with nomenclature used in some of these works we will refer to the core of M94 specifically, as \m94.

We compare the results from fitting the \m94 SED with those previously obtained for BHBs and AGN. The BHBs span a wider range in jet properties and we will limit our comparison to the ranges previously found for the different spectral states. The AGN, however, merit an individual comparison: \m94 is the closest ``pure" LINER and has a dynamical mass of $6.77\pm1.54\times10^6$ M$_{\odot}$ \citep{KormendyHo2013}. M87 is also a type 2 LINER \citep{Ho+1997} and has a similar bolometric luminosity in Eddington. M81 has both LINER and Seyfert characteristics, while the narrow line Seyfert 1 (NLS1) \4051 is a poster child for Seyfert galaxies; it is one of the original six studied by \citet{Seyfert1943}. Due to its relative proximity \sgra is not classifiable as LINER or Seyfert, but it displays extremely low Eddington luminosities (\lbol/\ledd $\sim2\times10^{-6}$) and at 4$\times10^6$ \msol \citep{Gillessen+2009} its dynamical mass is very similar to that of \m94. 

We introduce the new \m94 \emerlin observations in Section \ref{sec:datrad} and explain which other data we add to these radio observations in Section \ref{sec:4736obs}. Details about the \citetalias{MarkoffNowakWilms2005} framework are in Section \ref{sec:model}, while the best-fit models are presented in Section \ref{sec:nresults}. We discuss these results in Section \ref{sec:ndisc} and end with our conclusions in Section \ref{sec:nconcl}.

%%%%%%%%%%%%%%%%%%%%%%%%%%%%%%%%%%%%%%%%%%%%
%%%%%%%%%%%%%%%%%%%%%%%%%%%%%%%%%%%%%%%%%%%%%%%%%
%DATA%%%%%%%%%%%%%%%%%%%%%%%%%%%%%%%%%%%%%%
%%%%%%%%%%%%%%%%%%%%%%%%%%%%%%%%%%%%%%%%%%%%%%%%

\section{New \emerlin and archival VLA observations}
\label{sec:datrad}

\begin{figure}
\centering
\includegraphics[width=.49\textwidth]{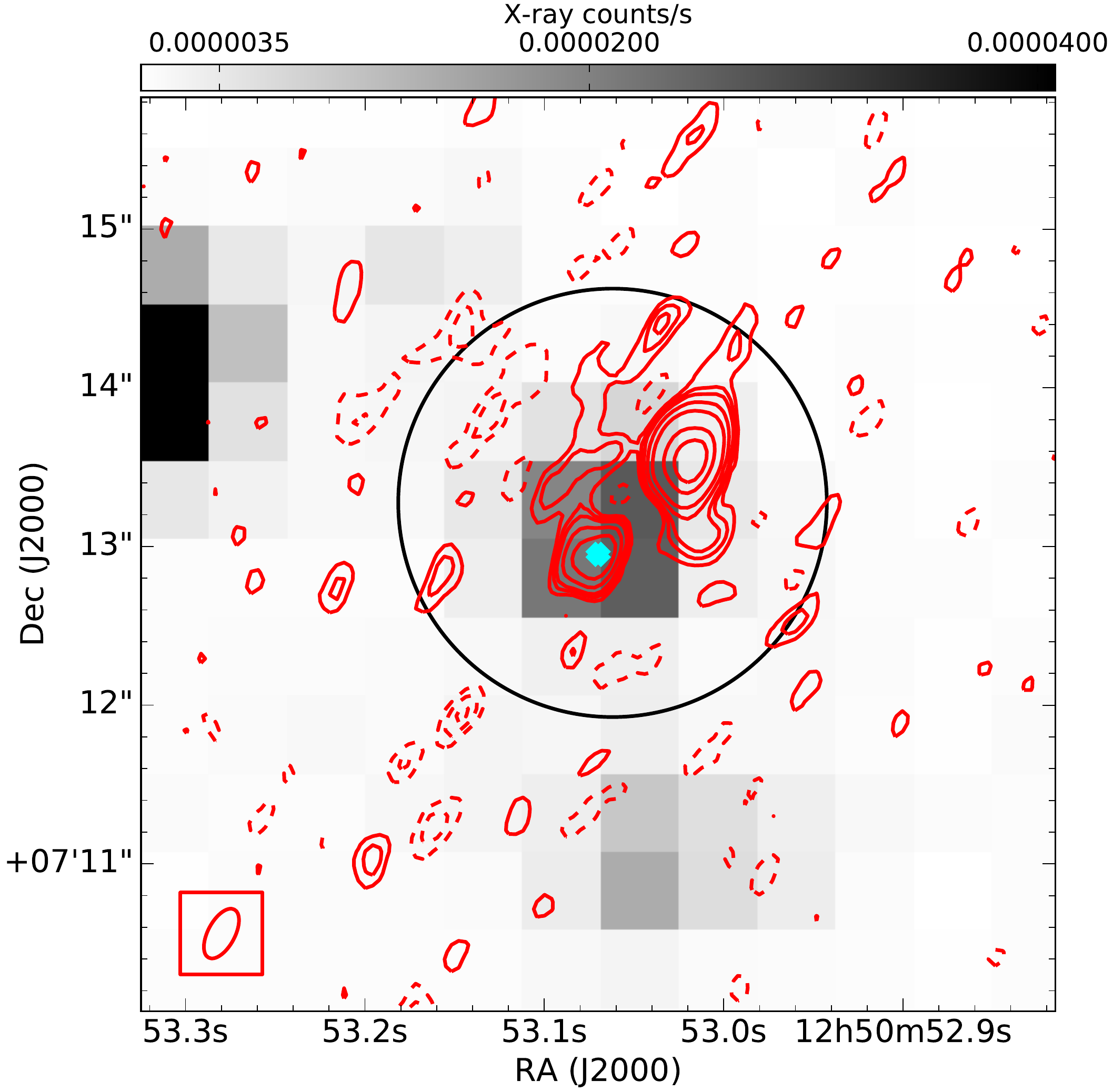}
 \caption{\emerlin M94 L-band contour map (in red) overlaid on the \chandra data from 2000 (greyscale; see Section \ref{sec:datxray}). The black circle denotes the region used to extract the \chandra spectrum, while the cyan dot denotes the location of the \m94 flat-spectrum nucleus at 15 GHz \citep{NagarFalckeWilson2005}. Radio contour levels are at -7, -5, -3, 3, 5, 7, 12, 20, 45 and 70 $\sigma$, where $\sigma=21$ $\mu$Jy (the latter two contour levels are only relevant for the extended source towards the north west). Dashed contours indicate negative levels. As there are negative levels down to $\sim$-5 $\sigma$, also levels up to  $\sim$ +5 $\sigma$ should be regarded with caution. Hence considering contours of greater than 7 $\sigma$ only, \m94 appears unresolved.}
 \label{fig:radmap}
\end{figure}

In this section we present a dataset obtained by the recently upgraded \emerlin array. These upgrades have enabled a dramatically increased bandwidth coverage, compared to the original MERLIN configuration \citep[e.g.][]{Garrington+2004}. These L-band ($\sim$1.5 GHz) observations are part of the Legacy \emerlin Multi-band Imaging of Nearby Galaxies Survey (LeMMINGS\footnote{\texttt{www.jb.man.ac.uk/$\sim$rbeswick/LeMMINGS/}}). Details about the reduction process are in Section \ref{sec:datradred}. This is the first time the double radio structure in M94 (seen at 8.49 GHz VLA; \citealt{KoerdingColbertFalcke2005}, but also at 4.9 GHz, see below) has been resolved into individual sources at low frequencies (see Figure \ref{fig:radmap}). The weaker of the two (at 1.5 GHz), towards the south-east, is unresolved within the noise and should therefore correspond to the nucleus, while the other source is extended and of unknown origin. We will investigate the nature of these two sources in more detail below.

\subsection{\emerlin radio reduction}
\label{sec:datradred}

All data inspection, flagging, calibration and imaging is done using the Astronomical Image Processing System (AIPS; \citealt{Greisen2003}), following the \emerlin cookbook\footnote{\texttt{www.e-merlin.ac.uk/data\_red/tools/e-merlin-cookbook\_V3.0\_Feb2015.pdf}}. M94 was observed by all 7 \emerlin stations (Lovell, Jodrell Mk2, Knockin, Defford, Pickmere, Darham and Cambridge) in L-band and in spectral line observing mode on May 5th and May 7th 2014. The phase calibrator is J1242+3751, the bright point source calibrator is 1407+284 (or OQ 208) and the absolute flux density calibrator is 1331+305 (3C 286). The array observed the target and phase calibrator alternately at a $\sim$7 min / $\sim$3 min cycle, respectively. The time on source totals $\sim13.7$ hours. The full bandwidth was 512 MHz, split into 8 spectral windows (SPW)s with each SPW split into 512 channels.

On the 5th of May no flux calibrator was observed, so all fluxes are calibrated using the 1331+305 observation of May 7th. This is usual practice with \emerlin and does not add significant uncertainty, as long as there are no changes in the instrumental system. 

Before imaging (with IMAGR), the target visibilities were re-weighted according to individual antenna sensitivity. The calibrated data were imaged with natural weighting to maximise sensitivity. Some iterations of self-calibration were attempted, but with a target flux below 1 mJy and an image dynamic range (DR) of only $\sim$200, the self-calibration did not lead to improvements. Multi-scale cleaning was attempted, but did not yield an improvement in noise level or DR. 

We use separate, source-free noise images, created using a shift in right ascension of 60\arcsec, to get an estimate of the RMS of the target image, rather than estimating the RMS from the target image itself. Using this method we can uniformly obtain reasonable estimates of the noise in any target image we make (see Section \ref{sec:extblob}), in a way that is less sensitive to local side-lobes\footnote{Such side-lobes can appear due to bad data that has gone unnoticed during flagging and/or the lack of self-calibration.} whose exact locus can vary from SPW to SPW. We note that despite the advantages of using an offset noise image, it may sometimes lead to an under-estimate of the true RMS, as it may fail to account for RMS contributions of the local side-lobes mentioned above. However, any such additional error is more systematic than statistical, and hence we prefer to use the errors obtained from the noise images for fitting purposes.

Using 100\,000 iterations for the cleaning process yielded images that portray the lowest RMS of $\sim$ 20$\mu$Jy/bm. Corresponding example radio contours are shown in Figure \ref{fig:radmap}.

\subsection{Extended steep-spectrum radio source}
\label{sec:extblob}

\begin{figure}
\centering
\includegraphics[width=.49\textwidth]{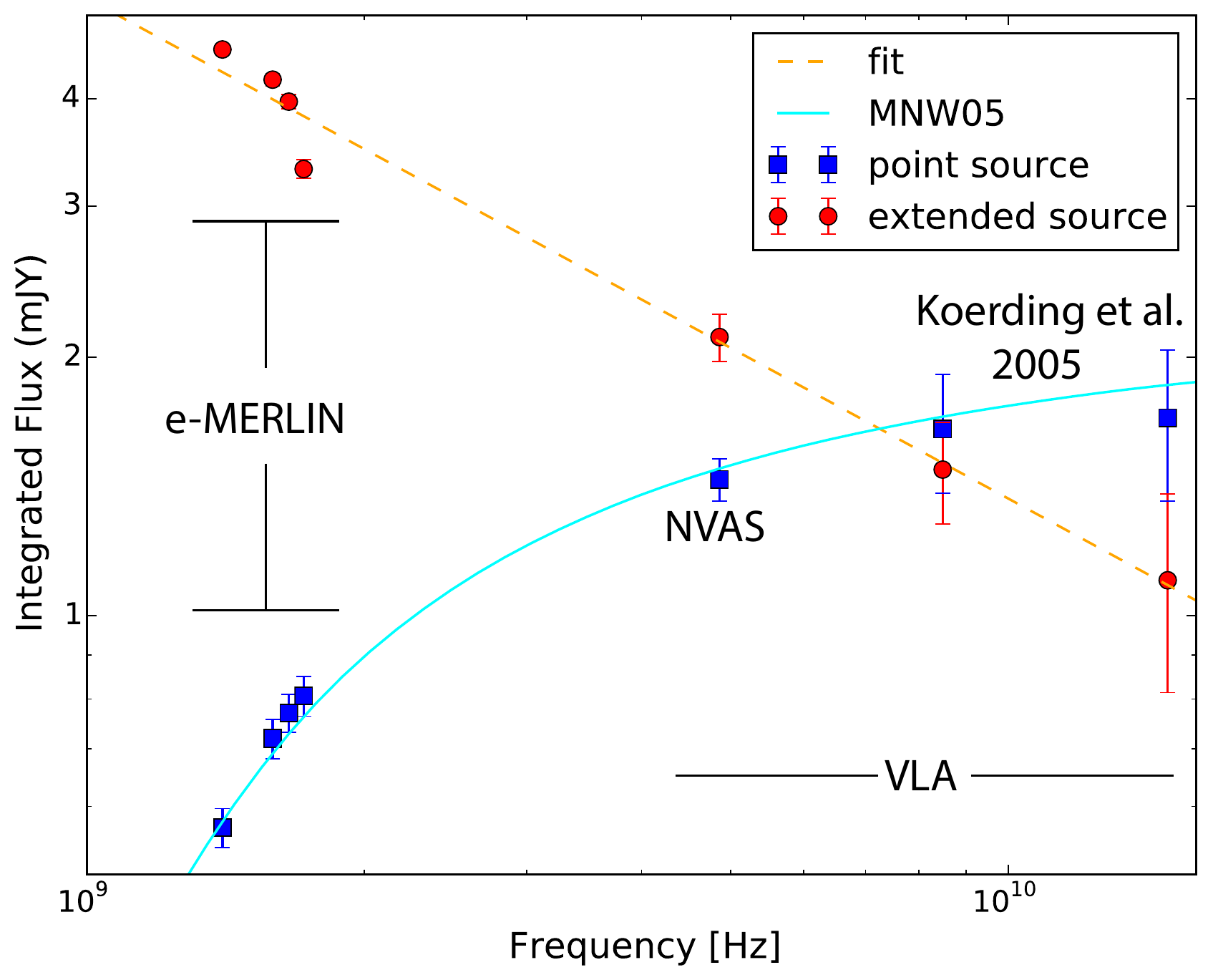}
 \caption{M94 1.5--15 GHz radio spectrum for the point- and extended sources in Figure \ref{fig:radmap}. The origin of these data are summarised in Table \ref{tab:mwdata}. The (unresolved) point source spectrum is shown with a jet model running through, while a power-law fit is applied to the (extended) non-core spectrum, best fit with a spectral index $\alpha=0.58$, where $S_{\nu}\propto\nu^{-\alpha}$.}
 \label{fig:radspec}
\end{figure}

Since this is the first time the double radio structure in M94 has been resolved in L-band and since the two components differ slightly in coordinates with those found by \citet{KoerdingColbertFalcke2005} it is prudent to investigate the two components, identify the nucleus, and ascertain the nature of the other component. To this end we take advantage of the \emerlin broadband capabilities and compile a detailed radio spectrum. Imaging all 8 available SPWs individually however revealed a poor data quality at the lowest-energy SPWs, due to RFI and possibly intrinsic absorption. Hence we bin together the 5 lowest energy SPWs but use individual images for the top three SPWs (cf. Figure \ref{fig:kntr}). We obtain integrated fluxes for both the point and extended source using AIPS/JMFIT. JMFIT reports the error on the fit, and these errors were added in quadrature to the RMS level of the separate noise images. The results are in Table \ref{tab:mwdata}, and plotted in Figure \ref{fig:radspec}.

In addition to the \emerlin observations, we have populated the radio SED with other VLA (A-configuration only) high-resolution C, X, U and K band data points (at 4.9, 8.5, 15 and 22 GHz respectively).

The C and K band observations were obtained as pre-reduced images from the automated VLA pipeline NVAS (NRAO VLA Archive Survey)\footnote{\texttt{http://archive.nrao.edu/nvas/}}. The C band images, even though dated as far back as 1983 and 1985, already show the double structure mentioned above (however we found no publication regarding this finding). \m94 is not detected at K band in A-configuration, due to an RMS of 0.718 mJy that we will use to include a 3$\sigma$ upper limit in the broadband SED (Figure \ref{fig:sed}). As this 22 GHz data offers no extra information regarding the double structure we ignore it in figure \ref{fig:radspec}.

To improve S/N we average the five 8.49 GHz, A-configuration measurements from \citet{KoerdingColbertFalcke2005}. The observations were done over a 3 month period and although the absolute flux densities reported for both the point and extended source show some variation, within the errors both the flux densities remained essentially constant.

The 15 GHz measurement is reported in \citet{NagarFalckeWilson2002,NagarFalckeWilson2005}. These papers do not report a double structure, however at 0.34 mJy \citep{KoerdingColbertFalcke2005} the image RMS would have prevented them from detecting the extended source at a 5$\sigma$ level. \citet{KoerdingColbertFalcke2005} re-reduced the original data and confirmed the extended source at a 4$\sigma$ level, but accurate flux determination from this data remained impossible due to the high RMS. Hence they obtained new U band maps from which they report both peak and integrated nuclear flux densities of 1.7 mJy, matching the integrated flux in the Nagar et al. works. Koerding et al. also report a 1.1 mJy peak flux density for the extended source (and note that the 0.95 mJy integrated flux density probably has some flux missing due to the small integration box used for this source). Unfortunately they report neither errors on these fluxes nor the RMS for the new images, even though it is clear from context that the new U band images are less noisy than the Nagar et al. maps. Hence considering the extra uncertainty due to the ``missing" integrated extended source flux and to remain conservative for the point source flux we will use the Nagar et al. RMS of 0.34 mJy as a 1$\sigma$ error on the Koerding et al. 15 GHz fluxes in Figure \ref{fig:radspec}.

Comparing the spectral indices ($\alpha$; where $S_{\nu}\propto\nu^{-\alpha}$) in Figure \ref{fig:radspec}, we find that the point source in the south east has an inverted spectrum and hence unambiguously identify this source with the core. With $\alpha\propto0.58$ the north-western extended component appears to show steep spectrum emission, consistent with slopes seen in star forming regions ($0.5<\alpha<1.0$; e.g. \citealt{Condon1992,Dopita+2005,Clemens+2008}), or injection indices in extended synchrotron knots \citep[e.g.][]{Kardashev1962,Biretta+1991,Carilli+1991}. The spectral index found for this extended component by \citet{KoerdingColbertFalcke2005} (from 8.4 to 15 GHz) is $\alpha=0.42$: slightly less steep, but broadly consistent. Looking at the \emerlin point source spectrum separately, at around $-2<\alpha<-2.5$, the spectral index is quite inverted, which could be due to Synchrotron Self Absorption (SSA) or free-free absorption. Indeed, had the core been resolved the source would have not been compact enough for SSA to be a viable mechanism to explain the absorption, as the magnetic fields required would have been orders of magnitude too high to explain the observed brightness temperature.

As it seems we have correctly identified the core we correlate this source position with multi-wavelength data.

\section{Multi-wavelength observations}
\label{sec:4736obs}

\begin{figure}
\centering
\includegraphics[width=.49\textwidth]{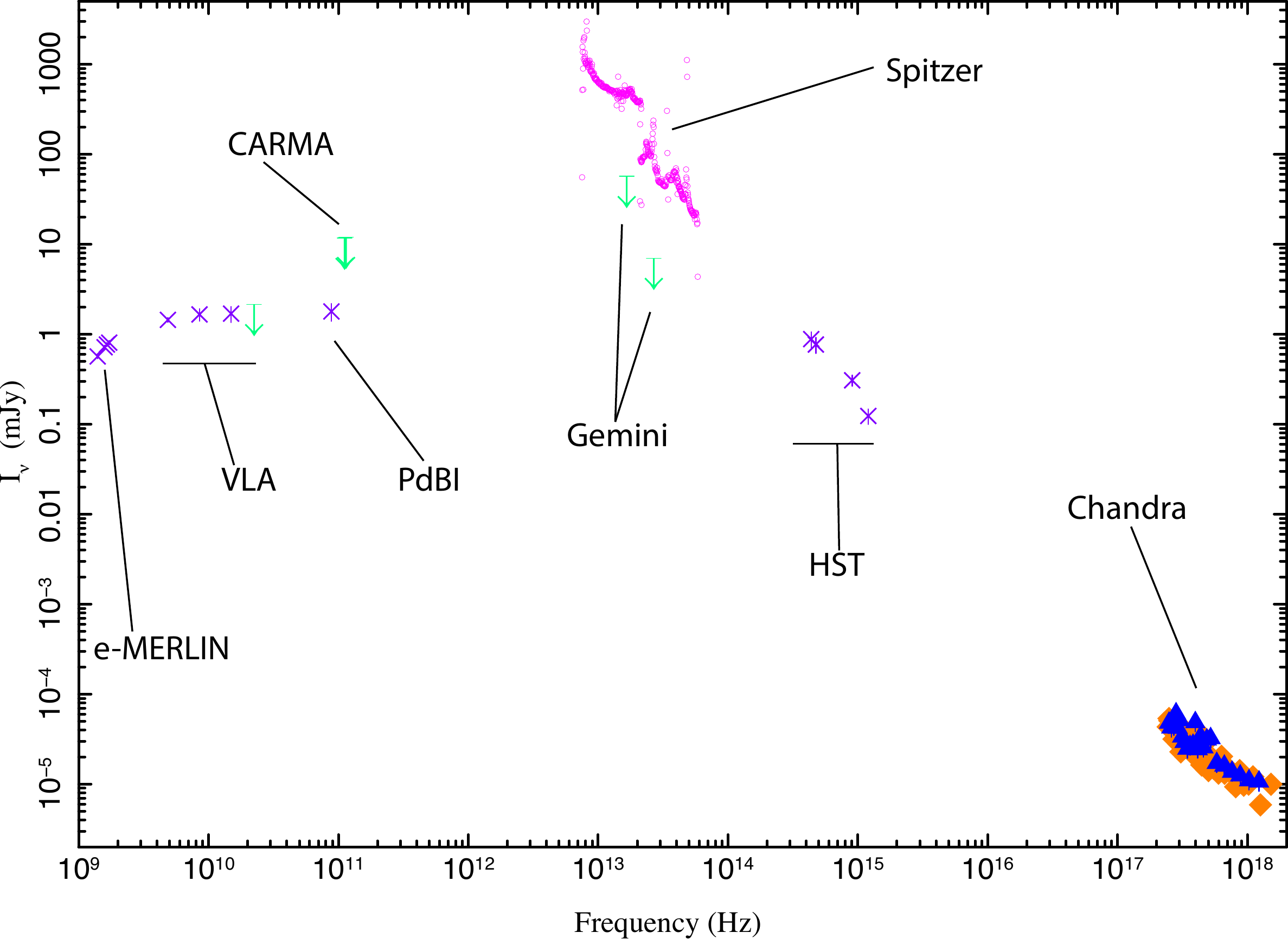}
 \caption{Multi-wavelength Spectral Energy Distribution (SED) of \m94, data only. For comparison we also include \spitzer data here (magenta open circles), which we will consequently ignore, as the two Gemini Michelle upper limits are clearly far better constraining. For details on all the data used, including precise observing frequencies, fluxes, expected resolutions, observation dates and an explanation of the upper limits, see Table \ref{tab:mwdata} and its caption.}
 \label{fig:sed}
\end{figure}

As M94 is relatively close, it is a well-studied source, observed by many different instruments. We are interested in \m94 and thus attempt to match the (0.15\arcsec) \emerlin resolution in the other wavebands in our selection of the available observations. The resulting SED (Figure \ref{fig:sed}) includes almost exclusively observations with high angular resolution (mostly sub-arcsecond, or $\lesssim$ 23 pc at the distance to M94, down to 0.1\arcsec, or 2.3 pc in optical). The exceptions to this high resolution are the millimetre observations, that are the only measurements available for this source, but were obtained in compact configurations, i.e. originally optimised for different science goals. However as we will explain below we are confident the millimetre data accurately represents the \m94 flux. Details of the observations are in the relevant subsections below, and are summarised in Table \ref{tab:mwdata}.

\begin{table*}\centering
\begin{tabular}{llccccrc}
\hline
Band				& Instrument			& Observed at		&Flux (mJy)		&Error (mJy)	& Scale (\arcsec)& Date			& Reference\\
\hline			
Radio			& \emerlin\	(L band)	& 1.40 GHz		& 0.57			& 0.030		&0.15		& 5/2014			& a\\
				& \emerlin\	(L band)	& 1.59 GHz		& 0.72			& 0.038		&0.15		& 5/2014			& a\\
				& \emerlin\	(L band)	& 1.65 GHz		& 0.77			& 0.039		&0.15		& 5/2014			& a\\
				& \emerlin\	(L band)	& 1.72 GHz		& 0.81			& 0.043		&0.15		& 5/2014			& a\\
				& VLA	(C band)		& 4.86 GHz		& 1.44			& 0.082 		&0.4			& 10/1983, 1/1985 & b \\
				& VLA	(X band)		& 8.49 GHz		& 1.65			& 0.26		& 0.24		& 6-10/2003		& c\\
				& VLA	(U band)		& 14.9 GHz		& 1.70			& 0.34		& 0.15		& 1/2001			& c, d\\
				& VLA	(K band)		& 22.4 GHz		& 2.15$^{\dagger}$	& 0.718*		& 0.96  		& 2/2000			& b\\
Millimetre			& PdBI				& 88.8 GHz		& 1.8				& 0.4			&5.7$\times$4.6& 8--9/2013 		& e\\
				& CARMA				& 111.5 GHz		& 11.1$^{\dagger}$	& 3.7*		&3.3$\times$2.3& 1/2007--3/2008	& a\\
				& CARMA				& 114.5 GHz		& 11.7$^{\dagger}$	& 3.9*		&3.2$\times$2.3& 1/2007--3/2008	& a\\
Infrared			& Gemini Michelle (Q band)& 18.1 $\mu$m	& 57$^{\dagger}$	& (19)		& 0.8			& 4/2007			& g\\
				& Gemini Michelle (N band)& 11.2 $\mu$m	& 13$^{\dagger}$	&  (3)		& 0.5			& 4/2007			& g\\
Optical			& \emph{HST} STIS		& 6867  \AA		& 0.881			& 0.176		& 0.1			& 6/2002			& h \\
				& \emph{HST} STIS		& 6295 \AA		& 0.768			& 0.154		& 0.1			& 6/2002			& h \\
Ultraviolet			& \emph{HST} ACS F330W& 330 nm		& 3.1 			& 0.043		& 0.5			& 6/2003			& i\\
				& \emph{HST} ACS F250W& 250 nm		& 0.123 			& 0.024		& 0.5			& 6/2003			& i \\
X-ray			&\emph{Chandra}-ACIS	& 1 - 8 keV		& see Fig \ref{fig:xraysed}&		& 1.35		& 5/2000, 2/2009	& j\\
\hline
\end{tabular}
\caption{Summary of observations of the multi-wavelength data included in the nuclear SED in Fig \ref{fig:sed}. For completeness we quote in parentheses the Gemini Michelle 1$\sigma$ errors, however these are unused as we use the two data points as upper limits (see text). The other three upper limits used (VLA K band and CARMA) are 3$\sigma$ and the values quotes in the error column here denote the image RMS and are marked with an asterisk. The millimetre observations as well as the upper limits may include contributions from the extended source, described in section \ref{sec:extblob}. References: a) This paper, b) NRAO VLA Archive Survey, c)  \citet{KoerdingColbertFalcke2005}, d) \citet{NagarFalckeWilson2002,NagarFalckeWilson2005}, e) \citet{vanderLaan+2015}, g) \citet{Asmus+2014} , h)  \citet{Constantin+2012}, i) \citet{Maoz+2005,Eracleous+2010}, j) \citet{Eracleous+2002}.}
\label{tab:mwdata}
\raggedright
%$^1$\
$\dagger$ Upper limit.\\
* Image RMS.
\end{table*}

\subsection{Millimetre}
\label{sec:datmm}

At millimetre wavelengths we have observations from the Plateau de Bure Interferometer (PdBI)\footnote{Now known as the NOrthern Extended Millimeter Array (NOEMA), after a recent upgrade to $>6$ antennae.} at $\sim90$ GHz and the Combined Array for Research in Millimeter Astronomy (CARMA) at $\sim110$ GHz.

\subsubsection{PdBI}

The central region ($\sim130$\arcsec\ diameter) of M94 was observed with 5 PdBI antennae in its D configuration during autumn 2013 (PI Van der Laan). Observations were centred at 88.8GHz, with a 3.6GHz initial bandwidth at 2MHz resolution given the WIDEX correlator. Data reduction was done via the dedicated PdBI pipeline. The output was checked, and generally found to be excellent. Four out of six observational tracks included an absolute flux calibrator (MWC349 and/or LKHa101), and those tracks could be used to also flux calibrate the other two tracks. Overall the absolute flux error can be assumed to be less than 10\%. Finally, data quality limits were set to limit the influence of bad observations. This way $\sim10$--20\% of visibilities were dismissed. The observing time remaining after calibration and reduction totals 13.3 hours spread over 11 pointings nearly uniformly, giving 1.2 hours for the central pointing.

A continuum uv-table was constructed from the remaining WIDEX visibilities, where 160 MHz-wide frequency ranges covering HCN(1-0) and HCO+(1-0) line emission (at 88.54 GHz and 89.10 GHz, respectively) were excluded. This leaves a 3.3 GHz total bandwidth. The sensitivity across the full 3.3 GHz bandwidth is 0.065 mJy/beam. Strong emission was found at position RA 12h50m53.068s, Dec 41d07m12.88s, near the phase centre of the observations. This emission was investigated with the GILDAS* routine uv\_fit. The emission is unresolved (i.e. is a point-source) with a flux of 2.0 mJy. By measuring the flux in the uv-plane the fitting error remains small, and is only truly dominated by the absolute flux error.

Even though the resolution of the PdBI in D configuration is at least an order of magnitude lower than those of the measurements in the other wavebands (beam PSF is 5.7\arcsec\  by 4.6\arcsec), no other significant continuum signal was found within the central 40\arcsec\  radius. In addition the positional accuracy of the flux measurement is an order of magnitude higher than the PSF size, since it was performed in the UV-plane. Although this unresolved measurement may include some flux from the secondary source observed in radio, extrapolating the tightly correlated spectral index in Figure \ref{fig:radspec} this non-core source should contribute no more than $\sim$0.4 mJy at these frequencies. Due to the error in absolute flux measurement the true flux density would lie between 1.8--2.2 mJy, however the core only flux could in reality be $\sim$0.4 mJy lower. Hence we will use a flux density of 1.8$\pm0.4$ mJy for the PdBI measurement, so we can be sure this data point is a good representation of the core emission at this particular frequency, despite the larger PSF.

\subsubsection{CARMA}

The continuum observations at 111.5 GHz and 114.5 GHz presented in this paper have been obtained through the CArma and NObeyama Nearby galaxies (CANON) CO (1-0) Survey. In the CANON survey, data from the CARMA and Nobeyama Radio Observatory 45-meter (NRO45) single dish telescope are combined to image the centres of nearby spiral galaxies in the (J=1-0) transition of $^{12}$CO (Koda et al. in prep). Due to its higher sensitivity on the relevant spatial scales, here we present only the CARMA data for the centre of M94.

The CANON observations of M94 at CARMA were taken in eight executions in the C and D configurations from January 2007 through March 2008 and are described in more detail in \citet{DonovanMeyer+2013}. The survey was designed to provide continuous velocity coverage around the CO (1-0) line with high spectral resolution and used wide continuum windows for calibration. Here we present measurements from the wide continuum windows in both the upper and lower sidebands. The windows are each 406 MHz wide and are centred at 111.5 GHz and 114.5 GHz. The field of view of the CARMA 19-pointing mosaic is 2.3\arcmin, centred on the position of M94.

The reduction was done using the usual CANON calibration routines in miriad which were adjusted to calibrate the continuum windows rather than the spectral line windows. The bandpass calibrators were 3C273 and J0927+390, with phase calibrators J1310+323 and J1159+292. The visibility flagging was not excessive. 

The beam size in the 111.5 GHz image is 3.3\arcsec\ by 2.3\arcsec\ and the beam size in the 114.5 GHz image is 3.2\arcsec\ by 2.2\arcsec. The images were created by inverting the visibilities with robust=-2 (a weighting that gives the less-well-populated longer baselines equal weighting and so pushes for a smaller beam, at the cost of slightly higher RMS). The RMS values measured in the centre of the maps are 3.7 mJy at 111.5 GHz and 3.9 mJy at 114.5 GHz. There are no continuum sources in either image within the CARMA field of view.

\subsection{Infrared}
\label{sec:datir}

Although \spitzer IRS spectral data is available, at a slit width of 3.6\arcsec\ the resolution of this instrument is relatively coarse and sure to be dominated by starlight and/or dust in the IR. \citet{Asmus+2014} analysed archival data from the 8.1m ground-based Gemini telescopes, as part of the Sasmirala (Subarcsecond mid-infrared view of local active galactic nuclei) survey. The survey features $N$ and $Q$ band M94 observations, obtained in 2007 with the Michelle spectrograph, offering nuclear/continuum flux measurements at 12 and 18 $\mu$m. These flux measurements were obtained using the {\sc mirphot} IDL routine that fits 2D Gaussians + a constant to the compact nuclear component in an attempt to subtract non-nuclear emission. Compared to the \spitzer measurements, the Sasmirala data constrains the \m94 nuclear flux to a level $\sim$10 times lower. However Asmus et al. note that the central 100 pc of M94 is completely dominated by extended starlight and so, also considering the range of possible other origins of flux in this waveband (see Section \ref{sec:model}), we will use these data points as upper limits. 

It is worth noting that \cite{RichingsUttleyKording2011} do not find evidence for a point source in the near-IR (in the \emph{Hubble} F160W filter, at around 1.6 $\mu$m).

\subsection{Optical}
\label{sec:datopt}

The two optical measurements included represent a spectrum obtained with the \hubble Space Telescope Imaging Spectrograph (STIS). The data were reduced and reported by \citet{Constantin+2012} being sampled over a rectangular aperture of 0.25\arcsec\ by 0.1\arcsec. Considering all the possible origins of ``pollution" of the optical (and UV) by non-jet sources we do not fit the detailed spectrum reported in that work, but  use the spectrum end-points only, assuming 20\% errors, to get an indication of the normalisation at these energies. In any case we do not expect the jet to be dominant in the optical/UV and for these reasons we do not fit the optical/UV data with a jet, but with an accretion disk, in order to take care of this normalisation (see Section \ref{sec:model}).

\subsection{Ultraviolet}
\label{sec:datuv}

In UV we use data reported in \citet{Eracleous+2010} that includes data from the \hubble Advanced Camera for Surveys (ACS), which has been automatically reduced by the Space Telescope Science Institute (STScI) pipeline as well as aperture photometry that has been done by \citet{Maoz+2005}, within a 10 pixel (0.27\arcsec) radius centred on the unresolved point source. These UV data are also used for SED fitting in \citet{Nemmen+2014}. \citet{Eracleous+2010} employed two different methods to de-redden the data, a starburst extinction law \citep{Calzetti1994} and a maximum correction corresponding to the Milky Way law of \citet{Seaton1979}. For our fluxes we take the average of the latter and the undereddened flux and use these extremes to determine the error bar sizes.

\subsection{X-ray}
\label{sec:datxray}

\begin{figure}
\centering
\includegraphics[width=.49\textwidth]{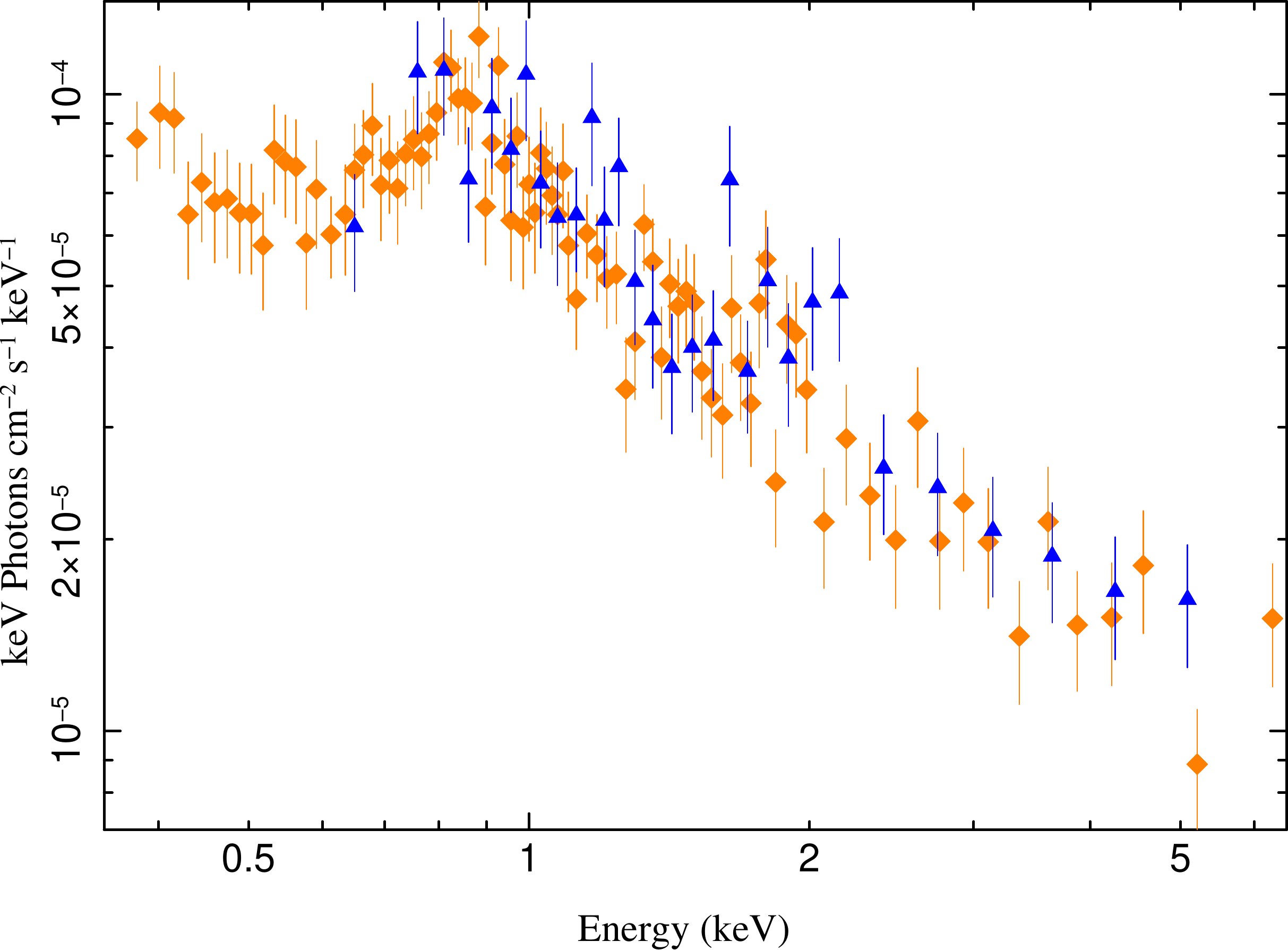}
 \caption{Background-subtracted X-ray SED of \m94, covering the full useable 0.36--8 keV \chandra range, binned to a minimum S/N of 4.5. The data was unfolded using the same model as used in Figures \ref{fig:syn}--\ref{fig:com}. We show the full X-ray range here for clarity, but only use the $>1$ keV data in this work (see text). Orange diamonds are Eracleous observations from 2000 while blue triangles are Jenkins observations from 2008.}
 \label{fig:xraysed}
\end{figure}

Using the Advanced CCD Imaging Spectrometer (ACIS), \chandra has observed M94 three times since its launch: twice in 2000 (ObsID 808, 47.37 ks and ObsID 410, 2.31 ks) and once in 2008 (ObsID 9553, 24.35 ks). Considering the relatively short exposure for ObsID 410 we choose to ignore this observation and reduce the other two using the standard CIAO (Chandra Interactive Analysis of Observations) software, v4.8 \citep{Fruscione+2006} and the CALDB version 4.7.2 calibration files. After extracting the spectra from the individual observations (with the CIAO tool {\sc specextract}) we fit for both datasets simultaneously, however the 2000/2008 observations have been performed in VFAINT/FAINT mode and ACIS-S/ACIS-I respectively. 

For both datasets we have to choose an appropriate extraction region. The normal pixel size of the CCD is 0.49\arcsec, while the PSF should be slightly smaller. We are interested in a point source, however the error in the absolute photometry is of the same order as the pixel size. In addition there is an energy dependence of the \chandra PSF such that the hard end of the PSF may have large wings that will fall beyond the extraction region if the latter is chosen too small. However the field is crowded, with close neighbouring sources limiting the size of the extraction region. Hence, we choose a circle that maximises surface area and includes the coordinates of the nucleus (at 15 GHz, which closely align with the PdBI observations, see above) but whose radius does not exceed the half-way points between the nucleus and the two closest neighbouring sources. The resulting extraction region is centred on coordinates 12:50:53.066 +41:07:12.73, has a radius of 1.35\arcsec\ (see Fig. \ref{fig:radmap}). To estimate the background we use 5 circular regions of 2.75\arcsec\ radius, randomly placed at empty regions roughly 1\arcmin\ away from the target. The final X-ray spectra are presented in Figure \ref{fig:xraysed}.

The usable energy range of \chandra is 0.36--8 keV. Below this range electronic noise dominates the signal from the sky X-rays, while above 8 keV there is negligible instrument response \citep{Doane+2004}. In Figure \ref{fig:xraysed} we see two components below 1 keV (that can be modelled well as a thermal plasma, at $\sim0.1$ and $\sim0.6$ keV, see \citealt{Roberts+1999,Pellegrini+2002}) contaminating the spectrum. These excesses observed in the X-ray SED seem due to larger-scale emission\footnote{Unrelated to the extended radio emission in Section \ref{sec:extblob}.}, e.g. hot, collisionally excited ISM, iron L and/or Neon lines (at $\lesssim$1 keV) and oxygen lines at lower temperatures, likely to be due to star formation and perhaps only indirectly related to the AGN by shocks. For an AGN at such low accretion rates there should be no soft X-ray contribution from the accretion disk blackbody emission, so these lower energies can be safely ignored by our model.  Therefore we truncate the spectrum to consider only X-rays $>1$ keV.

\subsection{Variability}
\label{sec:var}

The data in our SED are non-simultaneous. As explained in the introduction, this need not be a problem, since AGN tend to go through their evolution on extremely long timescales. LINERs can however be highly variable on timescales of months to years (with typical peak-to-peak amplitudes from a few to 50\%), particularly in the UV and X-ray bands \citep[e.g.][]{Maoz+2005,Pian+2010,Younes+2011}. Therefore we will briefly discuss the variability in the different domains, starting at low energy.

In the the radio through millimetre we may observe small fluctuations in the jet, consistent with adiabatic expansion or from normal physical variations in the input power, jet structure or accretion rate, travelling away from the nucleus as ``wiggles" in the spectrum. \citet{Markoff+2008} found evidence for such ``waves of variability" in M81*. However we see a consistently inverted spectrum, that increases almost perfectly monotonically with energy, at least up to PdBI frequencies. Hence, it seems unlikely that the variability in this domain should be significant at the time scales we are looking at, even though the observations used are spread over the past three decades in time (perhaps indicating a remarkably steady jet).

In infrared and optical bands we do not expect to see much variation, since we will mainly see galaxy light here. In the UV however, variation is expected for AGN. For \m94, \citet{Maoz+2005} measured a variability of 5 \% and 1 \% in the \hubble F250W and F330W ACS bands respectively, confirming an AGN could be responsible for a significant contribution in the UV bands. In addition, between 1993 and 2003 these authors observe long term variability of a factor of 2.5 in the UV, also indicative of an AGN. We do not expect these variations to be due to the jet. Moreover, for our modelling, variability in the IR-to-UV can be mostly ignored: As the jet is not expected to be the dominant component here (see below for discussion) it will not tell us much about the physics we want to study in this work. As will be explained in Section \ref{sec:model} we model these energies with a thin accretion disc as the dominant component and so the jet model results will not be sensitive to variability in the UV.

The X-ray variability of M94* has been examined several times before using different instruments. \citet{Roberts+1999} found no evidence for long-term (multi-year) variability comparing \rosat and \asca data, constraining the amplitude of the variation of observations four years apart to $\lesssim30$ \%). Also \citet{Pellegrini+2002} found no evidence of 2--10 keV flux variations between \asca and \bepposax observations, separated by five years. These instruments may conceal variation, however, due to a larger ratio of diffuse-to-core emission expected to be included in their relatively coarse PSFs, compared to \chandra and \xmm. Our \chandra observations were obtained 8 years apart. However integrating and comparing the 1-6 keV X-ray flux of these individual observations, we only see evidence for $\sim$2\% long term variability. \citet{Eracleous+2002} also look into the short-term variability for their dataset and constrained the excess variance to 0.06$\pm0.04$ s$^{-1}$, i.e at the 5--10\% level only at timescales of a few hours. 

From the above considerations we believe the variability in \m94 is insufficient to significantly influence our conclusions, as evidence suggests that it is a remarkably stable source. Hence the results from fitting the SED (Section \ref{sec:nresults}) should be robust enough to interpret and compare with other sources studied with \citetalias{MarkoffNowakWilms2005}.

%%%%%%%%%%%%%%%%%%%%%%%%%%%%%%%%%%%%%%%%%%%%
%%%%%%%%%%%%%%%%%%%%%%%%%%%%%%%%%%%%%%%%%%%%%%%%%
%MODEL%%%%%%%%%%%%%%%%%%%%%%%%%%%%%%%%%%%%%%
%%%%%%%%%%%%%%%%%%%%%%%%%%%%%%%%%%%%%%%%%%%%%%%%

\section{Jet paradigm}
\label{sec:model}

For spectral fitting we use the Interactive Spectral Interpretation System \texttt{ISIS} \citep{HouckDenicola2000}, compiled with \texttt{XSPEC} \citep{Arnaud1996} 12.8.2 libraries. \texttt{ISIS} allows us to forward-fold the model flux predicted by \citetalias{MarkoffNowakWilms2005} through the \chandra detector response matrices during fitting (while all non-\chandra data is fitted directly). 

For details of the \citetalias{MarkoffNowakWilms2005} model we refer to previous works of e.g. \citet{MarkoffNowakWilms2005,MaitraMarkoffBrocksopp2009,vanOers+2010}, but we will briefly describe the main parameters here. 

There are several physical parameters to the model that are fixed by observations, such as the BH mass (\mbh), the distance $d$ and jet inclination $i$. The jet itself is similar to a \citet{BlandfordKonigl1979} scenario, compact and stratified, consisting of multiple self-absorbed zones and has a length \zmax. All length scales are in units of gravitational radii (r$_g=GM_{\rm BH}/c^2$). The base of the jet, or jet ``nozzle", is cylindrical, with radius \rzero and height \hratio. The power entering the jets is represented by \njet, which is the dominant fitted parameter. \njet scales with the accretion power. The energy fed into the nozzle is divided between the kinetic and internal pressures, assuming the electrons are mainly responsible for the radiation, while cold (non-relativistic) protons carry the bulk kinetic energy. The ratio of the magnetic and particle energy densities is referred to as the equipartition factor $k=U_B/U_e$, with $U_B = B^2/8\pi$ the magnetic energy density, $B$ the magnetic field, and $U_e$ the electron energy density. The nozzle electrons are assumed to be in a quasi-thermal, mildly relativistic Maxwell-J\"{u}ttner velocity distribution of temperature \te, but at some length \zacc from the jet-base we assume 60\% of the electrons are accelerated into a power-law tail ($N(E)\propto E^{-p}$, where $p$ is a free parameter), assuming a process like diffusive shock acceleration (e.g. \citealt{Jokipii1987}). However we remain agnostic about the true acceleration process and parametrise the acceleration efficiency with a factor \escat that we fit for and in doing so absorb the unknowns. In all segments beyond the nozzle the jet is assumed to expand quasi-isothermally (see Crumley et al., in prep.), causing a longitudinal pressure gradient that accelerates the bulk-plasma asymptotically to Lorentz factors of 2-3 (derived from the Euler equations; \citealt{Falcke1996}). 

The \citetalias{MarkoffNowakWilms2005} framework incorporates cooling due to expansion, and synchrotron and IC losses under the assumption of a constant injection of fresh, high energy power-law leptons in each segment after the acceleration front. However if the cooling is fast, the injection rate may not be high enough to compensate for the cooling. If the dynamical timescale $t_{\rm dyn}$ of a segment (i.e. the lepton residence time) is at least of the order of the lepton synchrotron lifetime $t_{\rm syn}$, synchrotron losses will be significant. As a consequence the index of a steady state power-law electron distribution will be steepened by unity, from $p$ to $p+1$. The turnover point where the indices steepen is referred to as a \emph{cooling break} and should occur around $E_{\rm br}=\frac{24\pi m^2c^3}{\sigma_T B^2 t_{\rm dyn}}$ \citep{Kardashev1962}. 

We also multiply the outflow model by an absorption model (\texttt{phabs} in \texttt{ISIS}), parametrised by the line-of-sight-hydrogen density \nh\ (see Sec. \ref{sec:fixed} for details), which shows the biggest contribution in the softest X-rays.

We also include an optically thick, geometrically thin multi-colour accretion disk with a temperature profile of $T\propto R^{-3/4}$, following \citet{ShakuraSunyaev1973,Mitsuda+1984,Makishima1986}. This disk is parametrised by an inner radius \rin (with a temperature \tin) and an outer radius \rout. Although the photons coming from this disk are available for Compton up-scattering by the leptons in the base of the jet, for \m94 the contribution of this effect to the total model flux is insignificant. Furthermore, for LINERs the canonical ``big blue bump" (e.g. \citealt{Elvis+1994}) is often missing \citep[e.g.][]{Ho1999,Eracleous+2010}, hinting at a truncation/disappearance of a thin accretion disk. For LINERs the UV spectrum in $\log\nu$--$\log(\nu F_{\nu})$ representation is usually of comparable amplitude to the X-ray spectrum \citep[see e.g.][]{Fernandez-Ontiveros+2012}, however \m94 is $\sim100$ times brighter in UV than in X-ray. This is understandable as for most galaxies the flux in the IR-to-UV bands is expected to be an amalgamation of both galactic (e.g. stellar population and dust) and nuclear contributions (e.g. jet synchrotron, thermal radiation from a minor accretion disk, as well as reprocessed radiation from dust in a torus or narrow line region). In theory it should be possible to use a combination of stellar population synthesis templates (e.g. \citealt{BruzualCharlot2003}) and other models to account for all these components accurately. However the available data is insufficient to accurately disentangle the exact flux contributions in this energy range. Hence we make use of the (quasi-)thermal thin disk to model these 2 Gemini upper limits and 4 \hubble data points and account for the normalisation at this energy. In effect this disk subsumes all unseparated non-jet contributions from all the components mentioned above (stars, dust, etc). This approach allows us to focus on the broadband spectrum and implied physics of the compact jet source only. 

In the X-ray band, to account for the different \chandra modes/instruments/varying PSFs as well as possible source variability we allow a constant to determine the relative normalisation for the 2008 data compared to the 2000 dataset, as the latter has significantly more counts due to the longer observing time. Fitting data of different epochs should have the added benefit of further averaging out some of the expected variability in the X-rays (which is small to begin with, see Sec. \ref{sec:var}).

\subsection{Fixed and limited input parameters}
\label{sec:fixed}

\begin{table}\centering
\begin{tabular}{lccc}
\hline
Parameter			& Value					& Unit 		& Reference\\
\hline			
$d$				& 4.8 					& Mpc		& a	\\
\mbh				& 6.77					& $10^6$ \msol	& b	\\
$i$				& 29						& $^{\circ}$	& c	\\
\nh\				& 1.41--3.3$\times10^{20}$	& cm$^{-2}$	& d	\\
\hline
\end{tabular}
\caption{Summary of observable parameter values, and references: a) \citet{Eracleous+2010}; b)  \citet{KormendyHo2013}; c) \citet{Erwin2004,Trujillo+2009}; d) \citet{Stark+1992,Eracleous+2002}.}
\label{tab:obsval}
\end{table}      

As mentioned above we fix the observable model parameters, or put limits on the ranges allowed using results from the literature. These values and ranges are listed in Table \ref{tab:obsval}. 

\citet{Tonry+2001} determined the distance $d$ to M94 from surface brightness fluctuations. We however use a corrected value reported in \citet{Eracleous+2010}, who subtracted 0.16 mag from the distance modulus in Tonry et al., following \citet{Jensen+2003}. 
 
The black hole mass \mbh is a dynamical measurement quoted from \citet{KormendyHo2013}.

The compact jets in M94 have not been resolved and therefore the jet inclination is unknown. Moreover there is no correlation expected between the AGN jet inclination and the galaxy disk inclination \citep{Kinney+2000,Netzer2015}. The jet inclination could however be causally linked to the inclination of the inner galaxy disk. Based on the shape of the inner ovally distorted disk \citet{Erwin2004} reported a galaxy inclination of $i=35^{\circ}$. Consistent with this value,  \citet{deBlok+2008} found the galaxy disk inclination decreasing from $\sim50$--$29^{\circ}$ from the outer disk to the inner disk, based on \ion{H}{i} measurements. In our fits, lower inclinations are preferred to accommodate the relative flatness of the radio spectrum (higher inclinations yield a more inverted spectrum due to mild relativistic beaming). Hence for fitting we will use the bottom/inner disk value from \citet{deBlok+2008} of $i=29^{\circ}$. 

We limit the range for the absorption column \nh, requiring it to fall between the value corresponding to Galactic absorption alone \citep{Stark+1992} and the upper limit that was found by \citep{Eracleous+2002}, who fitted the 2002 \chandra spectrum data (that is also included in our SED) with an absorbed power-law. 

With regards to the electrons we make a few assumptions: We use a particle distribution index $2<p<2.5$, consistent with diffusive shock acceleration (e.g. \citealt{HeavensDrury1988,LemoinePelletier2003,Nemmen+2014} and references there-in). Also, as we see no evidence in the SED of a high-energy, exponential cutoff, we will assume the acceleration of the leptons occurs efficiently, and set \escat=0.01 in synchrotron dominated fits. This value is high enough to set the cutoff beyond the energy range covered by our \m94 data. Lastly this model is only valid for electron temperatures \te greater than $m_ec^2/k_{\rm B} = 5.94\times10^9$ K, as below this value the electrons would mainly emit cyclotron radiation instead of synchrotron radiation and we do not model the former separately.

%%%%%%%%%%%%%%%%%%%%%%%%%%%%%%%%%%%%%%%%%%%%
%%%%%%%%%%%%%%%%%%%%%%%%%%%%%%%%%%%%%%%%%%%%%%%%%
%RESULTS%%%%%%%%%%%%%%%%%%%%%%%%%%%%%%%%%%%%%%
%%%%%%%%%%%%%%%%%%%%%%%%%%%%%%%%%%%%%%%%%%%%%%%%

\section{Results}
\label{sec:nresults}

\begin{figure}
\centering
\syn
\includegraphics[width=.49\textwidth]{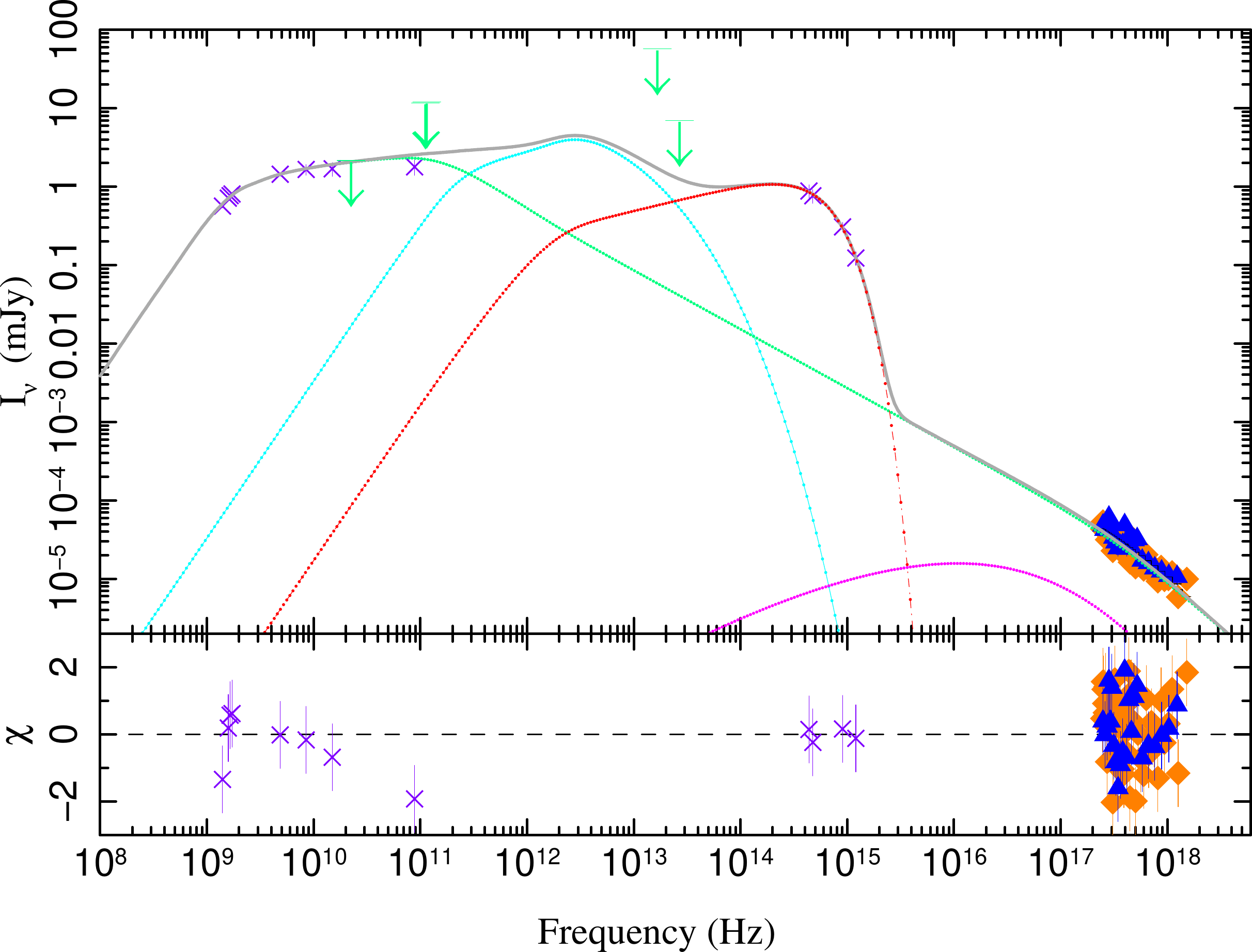}
\includegraphics[width=.49\textwidth]{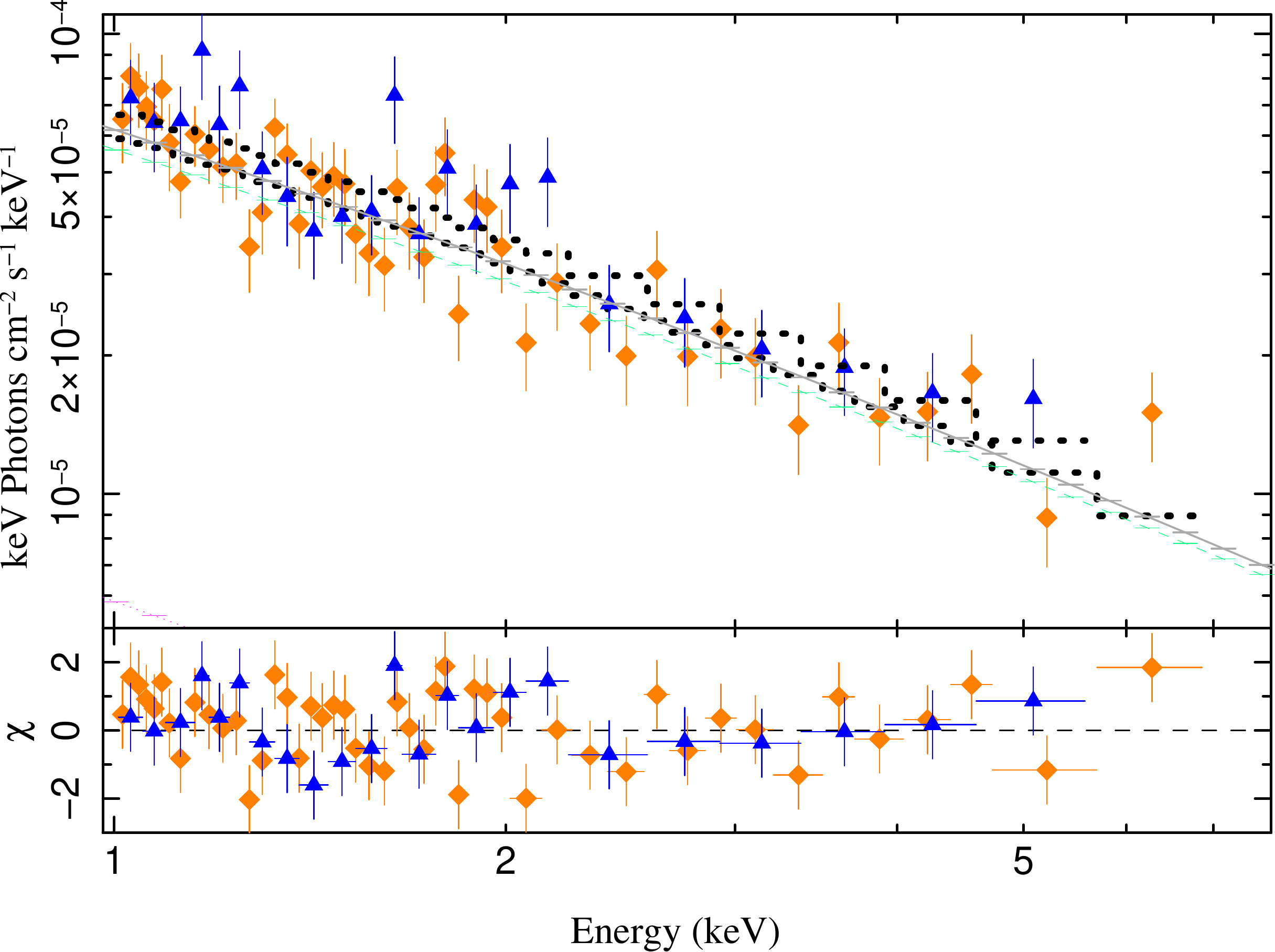}
 \caption{\m94 multi-wavelength broadband (top) and X-ray only (bottom) SED and best-fit model for the \syn fit (see Table \ref{tab:fitval}). The broadband SED is as in Figure \ref{fig:sed} (without the \spitzer data). The bottom plot is a simple magnification of the upper plot, but with the energy scale in keV instead of frequency. Hence all lines, colours and symbols are the same in both plots and will be defined below. Individual components from the \citetalias{MarkoffNowakWilms2005} model are shown with the following lines/colours: The cyan dotted curve is the jet pre-shock synchrotron contribution. The green solid curve is the post-shock synchrotron. The red dash-dotted curve represents all the contributions in the IR-to-UV, modelled by a multi-colour accretion disc. The magenta solid curve represents the Compton-up-scattered seed photons from both the synchrotron self-Comptonisation (SSC) of the pre-shock synchrotron and the accretion disc, although, considering the relative normalisations of those components, the latter only contributes weakly in these models. The solid grey line indicates the total of the above components (however not forward-folded through the detector response matrices). Only visible in the bottom plot, the black dashed lines show the forward-folded model fluxes (albeit only where data is present) for both X-ray data sets that are denoted by orange diamonds and blue triangles respectively (as explained in Fig \ref{fig:xraysed}).}
 \label{fig:syn}
\end{figure}

\begin{figure}
\centering
\syncom
\includegraphics[width=.49\textwidth]{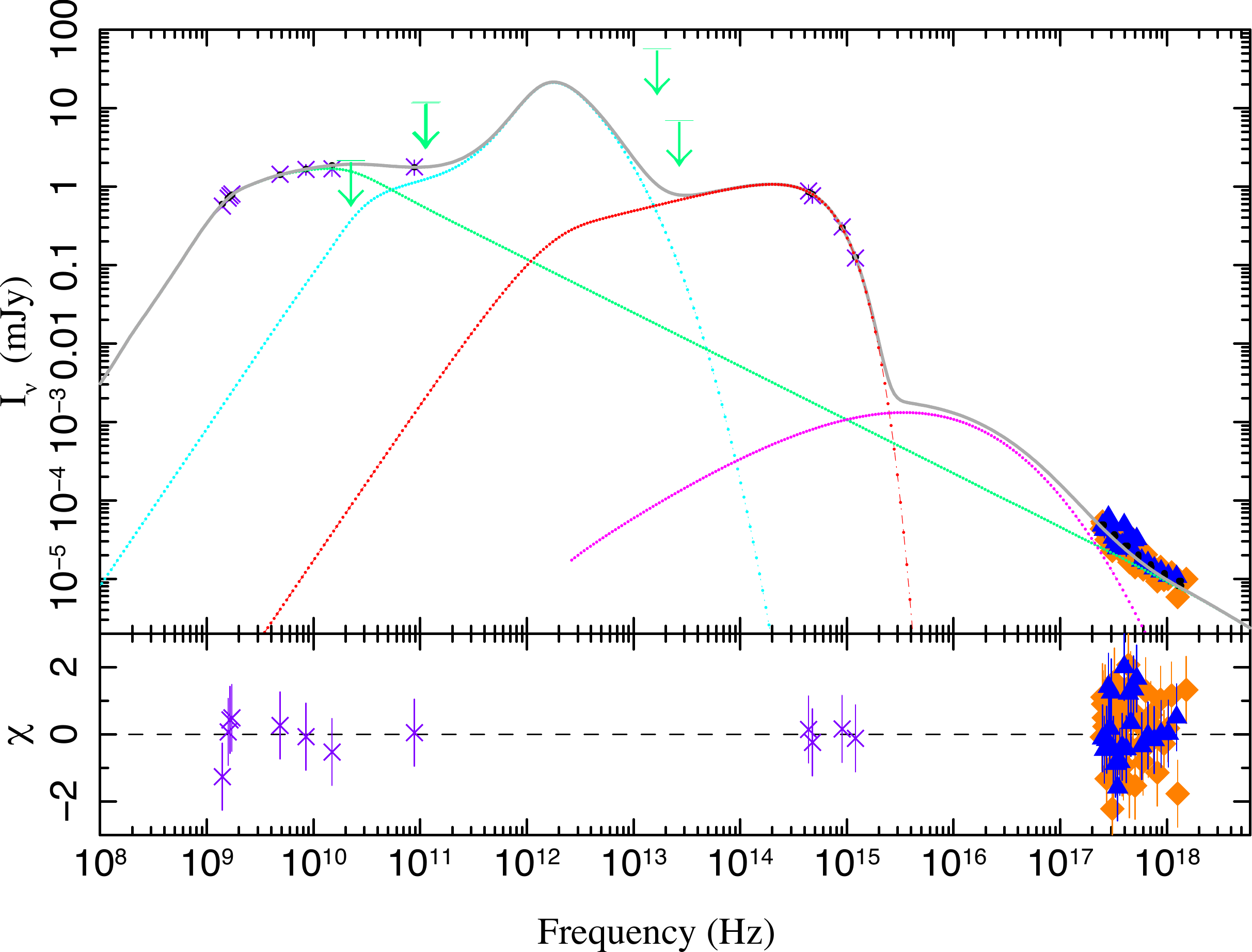}
\includegraphics[width=.49\textwidth]{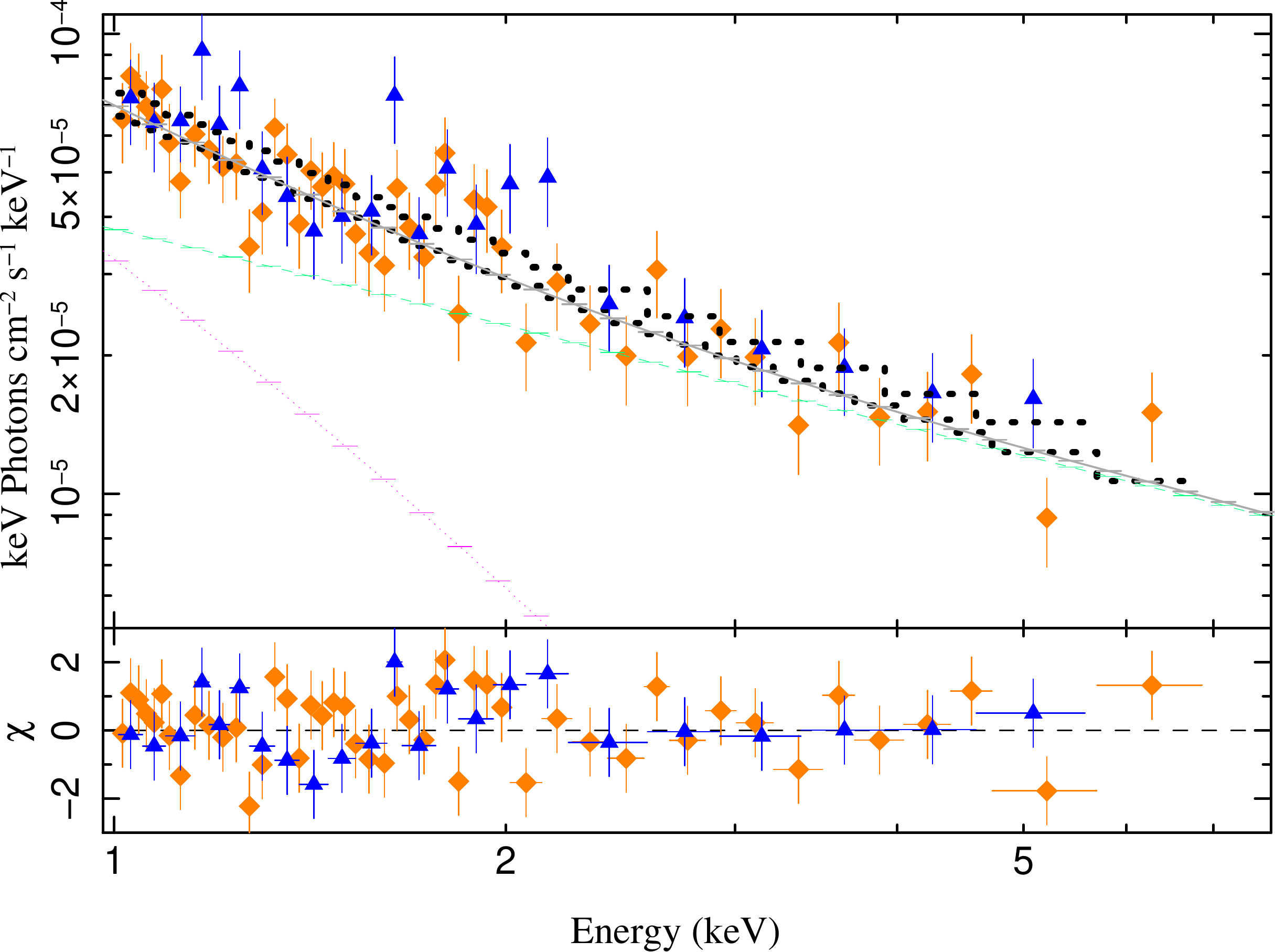}
 \caption{Multi-wavelength broadband (top) and X-ray only (bottom) SED and best-fit model of \m94 for the \syncom model (see Table \ref{tab:fitval}). Definitions of lines, colours and symbols are explained in Figure \ref{fig:syn}. Please note that in the top panel the (insignificant) SSC component has not been plotted below $\sim2\times10^{12}$ Hz, due to numerical breakdown of the model.}
 \label{fig:syncom}
\end{figure}

\begin{figure}
\centering
\com
\includegraphics[width=.49\textwidth]{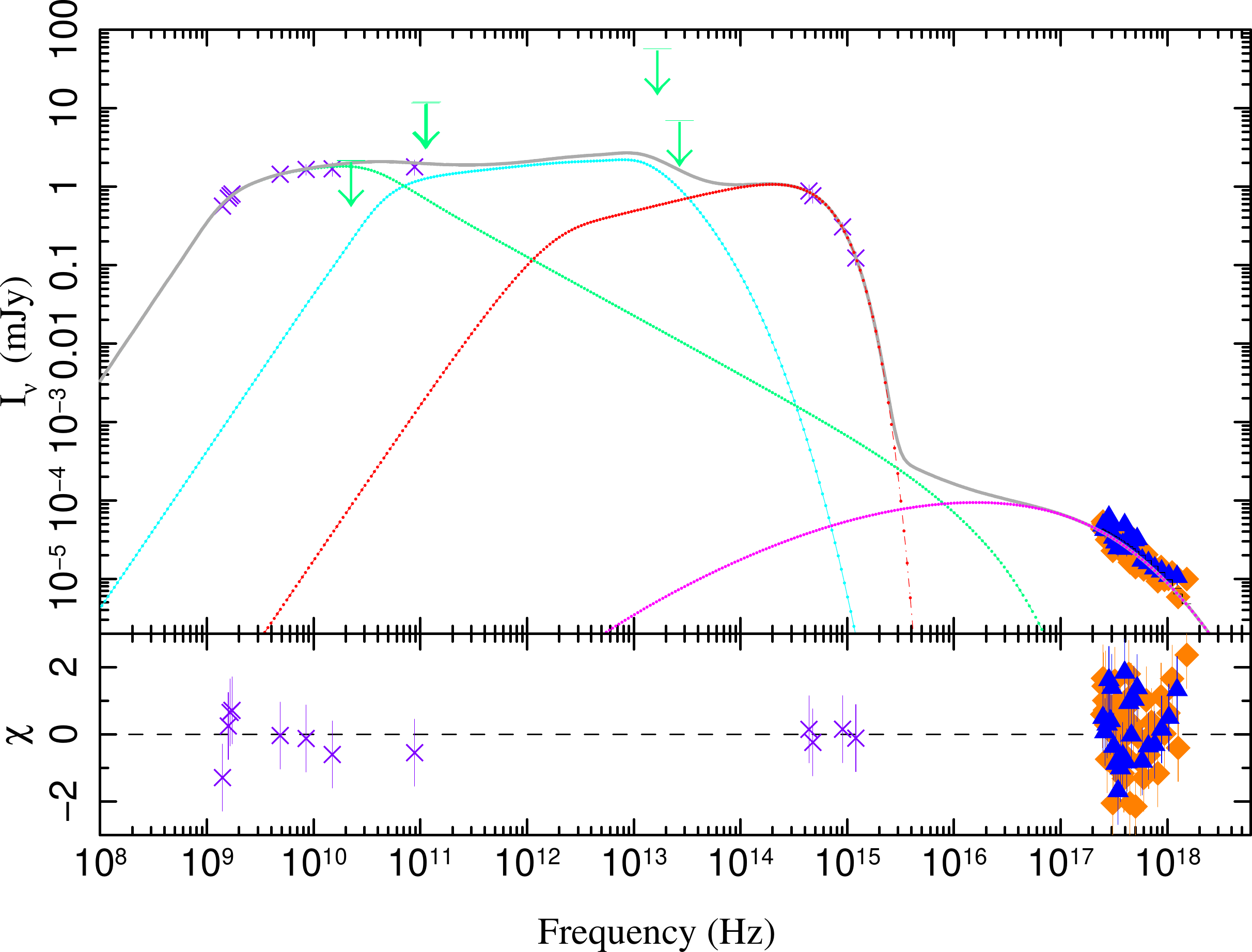}
\includegraphics[width=.49\textwidth]{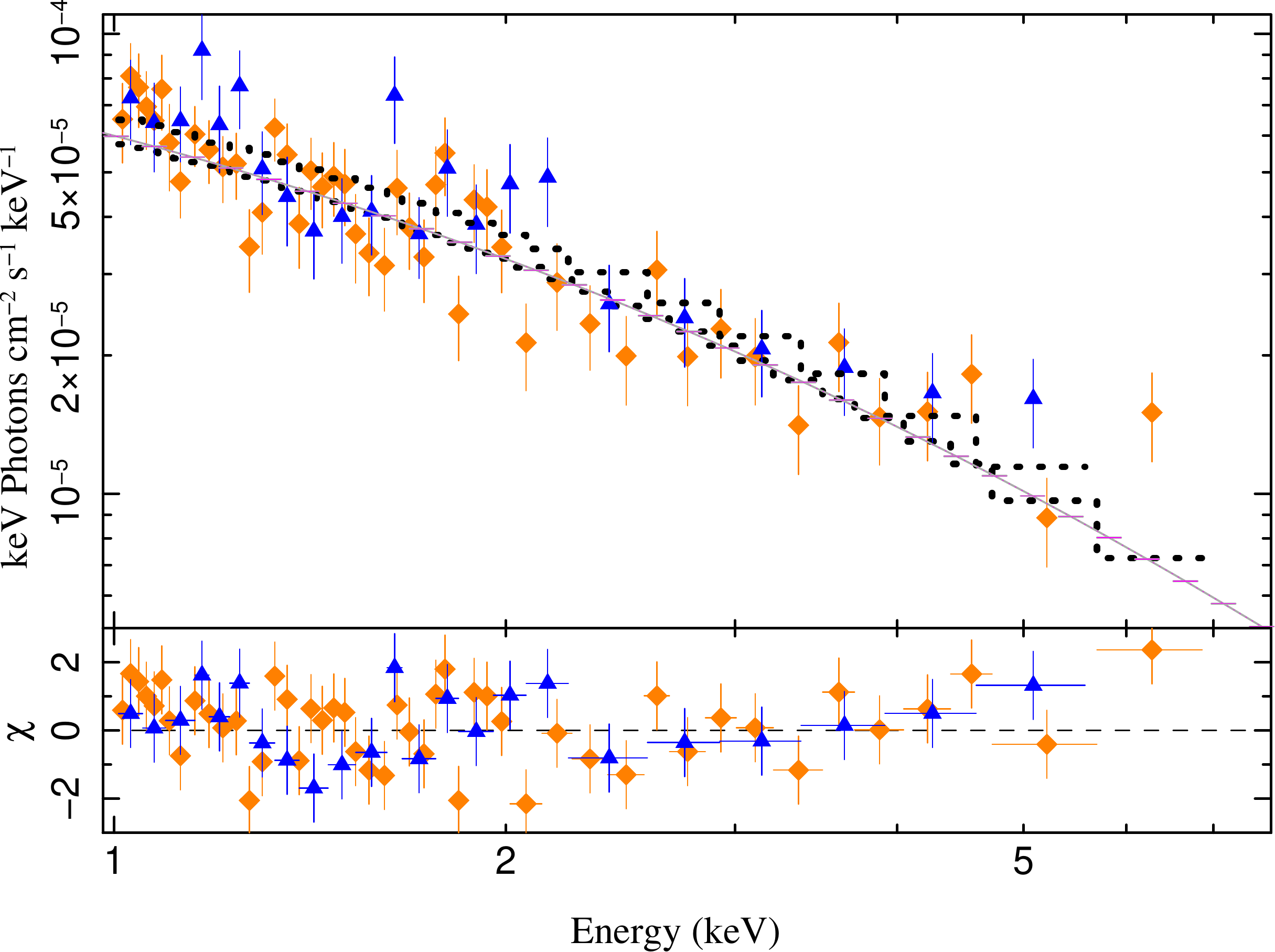}
 \caption{Multi-wavelength broadband (top) and X-ray only (bottom) SED and best-fit model of \m94 for the \com model (see Table \ref{tab:fitval}). Definitions of lines, colours and symbols are explained in Figure \ref{fig:syn}.}
 \label{fig:com}
\end{figure}

We use the {\sc isis} script \texttt{conf\_loop} to iteratively search for confidence intervals on free parameters, using a tolerance of $\Delta\chi=10^{-3}$. As our data were not obtained simultaneously, it is impossible to assess errors resulting from systematics due to possible variability in \m94. However, as explained in Section \ref{sec:var}, we see evidence of only limited variability in this source. Hence we include all the 90 \% confidence limits obtained\footnote{We do not quote confidence limits on the acceleration zone distance (\zacc) due to the segmented nature of the \citetalias{MarkoffNowakWilms2005} jet. As the jet is composed of discrete zones, neither the location nor the confidence limits can be accurately determined.}, in Table \ref{tab:fitval}. The associated fits are displayed in Fig. \ref{fig:syn} -- \ref{fig:com}.

{\renewcommand{\arraystretch}{1.3}
\begin{table}\centering\footnotesize	
\begin{tabular}{llrrr}

%\cline{4-5}
														\mc{5}{c}{\m94}						\\			
		
										&Parameter (Unit)				&\syn				&\syncom				&\com						\\
 \cline{1-5}                                                                                                                                                                                    
\mr{8}{*}{\rotatebox{90}{Jet\hspace{.15mm} }}		&\njet ($10^{-5}$\ledd)			& $6.8_{-5.0}^{+0.8}$	&$28_{-14}^{+51}$		& $35_{-18}^{+9}$ 				\\     
										&\te ($10^{11}$ K)				& $1.4_{-0.2}^{+0.3}$	&$1.0_{-0.1}^{+0.1}$		& $1.3_{-0.1}^{+0.2}$			\\          
										&\rzero (\rg)					& $3.7_{-1.7}^{+3.2}$	&$6.4_{-3.0}^{+3.1}$		& $1.0_{>-0.0}^{+1.1}$			\\
										&$k$							& $0.9_{-0.3}^{+10.9}$	&$0.10_{-0.05}^{+0.00}$	& $0.03_{-0.02}^{+0.11}$			\\
										&\escat						& 0.01*				& 0.01*				& $\mathit{8\times10^{-7}}$		\\
										&$p$							& $2.49_{-0.12}^{>+0.01}$&$2.44_{-0.03}^{+0.00}$	& $2.50_{-0.28}^{>+0.00}$		\\
										&\zacc (\rg)					& $\mathit{94}$			&$\mathit{768}$		& $\mathit{425}$ 				\\
										&\hratio						& 1.5*				&$12_{-4}^{+20}$		& $0.7_{-0.1}^{+2.8}$			\\
                                                                                                                                                                                                 
\cline{1-5}                                                                                                                                                                                     
\mr{3}{*}{\rotatebox{90}{Disk\hspace{.1mm} }}		&\rin (\rg)						&$37_{-8}^{+8}$		& $37_{-9}^{+9}$		& $36_{-10}^{+3}$            			\\
										&\tin ($10^3$ K)				&$10.6_{-1.2}^{+1.0}$	& $10.6_{-0.9}^{+1.1}$	& $10.6_{-1.0}^{+1.5}$			\\
										&\rout  (\rg)					&$10^4$*				& $10^4$*				& $10^4$*						\\				
\cline{1-5}                                                                                                                                                                                     
										& \multirow{2}{*}{\chisqdof}		& 78.4/73				& 68.7/72				& 81.2/72						\\
										&							&($\simeq1.07$)		& ($\simeq0.95$)		& ($\simeq1.13$)				\\

\end{tabular}
\caption{Best-fit model parameters for \m94 (see Section 4 for detailed description; The jet-related paramaters are: \njet = the scaled jet power, \te is the electron temperature in the jet base, \rzero is the radius of the jet base, $k$ is the ratio of thermal electron energy density to magnetic energy density, \escat relates to acceleration efficiency, $p$ is the accelerated particle distribution index, \zacc is the location where particle acceleration starts in the jets, \hratio is the scaled height of the jet base). An asterisk (*) indicates that a parameter was frozen at the value given. We failed to resolve both upper and lower limits of the error bars for parameters listed in italics, while we failed to resolve either an upper, or a lower bound for parameters with a greater-than sign ($>$, meaning the true confidence limit is farther removed from the parameter value than indicated by the limit value). To fit the \emerlin slope well at the lowest radio frequencies the length of the jet \zmax was fixed to $10^{16.15}$ cm (to avoid over-interpreting the model we do not allow this parameter freedom to vary, see text).}  
\label{tab:fitval}
\end{table}      
}

For most LLAGN it is difficult to constrain the dominant radiative mechanism \citep[e.g.][]{Plotkin+2012}, due to degeneracies within the possible radiative mechanisms. Firstly, within synchrotron there is an intrinsic degeneracy, where energy distributed into either the particle energy density or the magnetic energy density can work to enhance the spectral flux, since both an increase in lepton density (number of particles radiating), as well as a stronger magnetic confinement of these particles (individual particles radiating more) can increase the net flux. Secondly, both synchrotron and inverse-Compton processes can create power-law radiation. Hence, unsurprisingly, we find three statistically good models with acceptable physical model parameters. In the first fit (from here on referred to as \syn) the X-ray spectrum is well-approximated by optically thin synchrotron photons originating from the acceleration region at \zacc$\sim100$ \rg. The second fit has incrementally better statistics compared to the other two, and since the most salient difference is a significant synchrotron self-Compton (SSC) contribution in the softer X-rays we will refer refer to this fit as \syncom. Thirdly we obtained a fit where the X-rays are fully dominated by SSC (hereafter: \com). In the following we proceed to describe and compare these fits and explain why the \syn fit seems the most physically tenable. Consequently we will compare our \m94 modelling with other sources previously modelled with \citetalias{MarkoffNowakWilms2005} in Section \ref{sec:disccomp}.

\subsubsection*{\syn} The X-rays of the \syn fit are dominated by optically thin cooled synchrotron radiation, displaying a steepening of slope in the soft X-rays. Observing this break is consistent with findings in \citet{Russell+2013} who predict such a break between the UV and the X-rays at the X-ray luminosities (in Eddington) that \m94 portrays. The fit shows all confidence limits resolved, except for the upper limit of the particle distribution index $p$ (within the range allowed). A marginal SSC contribution is seen in the UV -- soft X-rays, however this component is not needed to adequately fit the data and hence the jet height-to-base ratio \hratio was fixed to 1.5, consistent with previous works. At $k\sim0.9$ the jet would be slightly sub-equipartition, however within the errors a magnetically dominated jet cannot be ruled out. Considering the large upper limit on the jet base radius, the nozzle appears less compact than in the other two jets.

\subsubsection*{\syncom} In absolute terms the \syncom model shows the best statistics of all three fits. The improvement in \redchisq is mainly due to a lower residual for the submm PdBI data point due to a slightly flatter model radio spectrum. In the \citetalias{MarkoffNowakWilms2005} model, mainly the inclination determines how flat or inverted the radio spectrum is, with higher inclinations leading to more inverted radio spectra. However the transition from an optically thick to an optically thin dominated spectrum can locally flatten the radio spectrum somewhat at an energy just above this transition (compare the top panels of Figure \ref{fig:syncom} with those of Figure \ref{fig:syn} and \ref{fig:com}; in contrast to the latter two Figures, the grey line that denotes the total model flux is not inverted above $\sim10^{11}$ Hz in Figure \ref{fig:syncom}). In the \syncom fit, this transition occurs at an energy just below the PdBI observing frequency, tailoring the radio/submm model spectrum almost exactly to the data. The X-rays, on the other hand, do not show improved residuals at all. We consider the transition effect to be a model artefact and the consequent improved statistics are thus not due to the model better describing the underlying physical picture. Therefore we find it likely that this fit represents a local or ``false" minimum. In addition, compared to the other two fits an extra SSC component is needed to fit the X-rays equally well. Using an extra fitting-component while not improving the actual model should generally be avoided to minimise the available degrees of freedom and avoid degeneracy. Hence we will exclude the \syncom fit from detailed further analysis, but have included it here for completeness. 

\subsubsection*{\com} As is also clear from the \syncom fit above, a significant SSC contribution requires an increasingly compact nozzle. In the \com fit this is evident from the relatively small jet base radius that pushes the lower allowed value of \rzero$=1$ \rg. As this value is within the 6 \rg innermost stable orbit of a Schwarzschild BH it suggests \m94 is rapidly rotating. An even smaller value of \rzero would place the entire nozzle within the innermost stable orbit of even a maximally spinning ``Kerr" black hole. The low value of the ratio of magnetic to particle distribution energies $k$ may however pose a bigger problem. Although values below unity could be possible, it is currently not quite clear what it means in the context of an \citetalias{MarkoffNowakWilms2005} jet, that is assumed to be maximally dominated. Due to uncertainties regarding energy redistribution in the jet nozzle it is difficult to quantify whether a sub-equipartition flow is physical or perhaps an artefact due to simplifications in the model (also see discussion in \citealt{Plotkin+2015}). In general, particularly for BHBs, the results are more consistent with equipartition. Thirdly, the acceleration efficiency parameter \escat needed to be fixed (within a truncated range $10^{-6}$--$10^{-8}$) to avoid the fit evolving towards a dominant optically thin synchrotron contribution in the X-rays (yielding a result similar to one of the two above). Consequently the optically thin synchrotron is not at all constrained by the X-rays. For the same reason the particle distribution index $p$ is not well-constrained. Although this fit seems to be pushing extremes, especially in terms of \rzero and $k$, we assert that within the uncertainties of the model this fit can not be excluded from consideration, particularly as this type of solution has been known to describe well other low luminosity sources like quiescent BHBs \citep{Plotkin+2015} and low luminosity AGN.

\subsubsection*{Further remarks on fits}

As explained in Section \ref{sec:model} we assume the presence of an accretion disk and use its quasi-thermal flux to account for the IR-to-UV flux measurements, although additional physical components are expected to contribute. The obtained accurate fit of the optical and UV \hubble data suggests a thermal origin (perhaps the high energy cutoff of a stellar population), but considering the uncertainties we draw no further conclusions from the obtained disk parameters\footnote{In all models the outer disk radius \rout was fixed to a typical value of $10^4$ \rg that avoids fitting any features of the SED with the low-energy part of the disk's thermal spectrum.}. We only list them in Table \ref{tab:fitval} for completeness.

As touched on above: Our fits to the radio spectrum suggest that the inclination chosen (29$^\circ$) could be on the high side. A lower inclination would yield a flatter radio spectrum and hence better accommodate the PdBI data point. We have set the inclination using observations from \citet{deBlok+2008}, who quote a value specific for the inner disk to which we assumed the jet \emph{may} be causally linked (emphasising again that the jet and galaxy disk inclination are not expected to be correlated). For completeness we note that \citet{Trujillo+2009} quotes a lower inclination of 8\deg, inferred from the ellipticity of 3.6 $\mu m$ isophotes. This value could be more appropriate for the innermost part of the source, since the \spitzer data used to infer this is of higher resolution and \ion{H}{I} as a tracer loses potency towards the nuclei of galaxies, as the gas becomes increasingly molecular. As expected, using the lower inclination value from Trujillo et al. resulted in (incrementally) better statistics for both synchrotron and SSC dominated fits, however the overall physical picture did not change. Hence we did not further pursue this avenue, but take this as tentative evidence suggesting the jet inclination in \m94 may be less than 29\deg.

We should note that in our fits the column density \nh was left free to vary in the range noted in Table \ref{tab:obsval}, but consistently pegged at the lower value of $1.41\times10^{20}$ cm$^{-2}$, reflecting Galactic absorption only. The upper limit of 3.3$\times10^{20}$ cm$^{-2}$ mentioned in Table \ref{tab:obsval} from \citet{Eracleous+2002} was obtained by modelling the X-ray spectrum with a simple power-law only model. As our modelling is more complex, and we only use data above 1 keV, the upper limit may not be relevant for our modelling. Additionally, in \citet{Roberts+1999} it is argued that any additional intrinsic absorption in \m94 may be masked by soft thermal emission. 

Aside from the difference in energy distribution in terms of magnetic to particle energy densities and acceleration efficiency, the two remaining fits \syn and \com actually reflect a rather similar physical picture as they agree on the remaining parameters within about a factor two. In fact not even the magnetic and particle densities are that different, as the \syn model only just exceeds equipartition while the \com model is not too far below. The electron temperatures, particle distribution indices and jet base-radius-to-height ratio are even in excellent agreement. So although we can not discriminate the dominant radiation mechanism, the underlying physical picture in terms of jet geometry and energy budget are well-represented by these two models, that we will compare below with previous explorations.

%%%%%%%%%%%%%%%%%%%%%%%%%%%%%%%%%%%%%%%%%%%%
%%%%%%%%%%%%%%%%%%%%%%%%%%%%%%%%%%%%%%%%%%%%%%%%%
%DISCUSSION%%%%%%%%%%%%%%%%%%%%%%%%%%%%%%%%%%%%%%
%%%%%%%%%%%%%%%%%%%%%%%%%%%%%%%%%%%%%%%%%%%%%%%%

\section{Discussion}
\label{sec:ndisc}

Recently \citet{Nemmen+2014} also modelled an \m94 broadband spectrum. However their modelling was not performed in the context of an outflow-dominated model, but rather focussed on an ADAF-dominant (Advection Dominated Accretion Flow; a type of RIAF) model. In fact, they preclude a jet from being able to fit the \m94 X-ray data, noting that the X-ray spectrum would be too hard for a diffusive shock acceleration origin. Their modelling, however, does not include the inverse Compton scattering of seed photons from the jet, while our results clearly indicate that seed photons from an outflow offer an abundant photon reservoir to feed a Compton component. In addition we show that a particle distribution index $p\sim2.4$--2.5 is adequate to interpret the data. Such values are fully consistent with diffusive shock acceleration (and consistent with \citealt{Nemmen+2014}). Hence this work clearly indicates that not only is a relativistic jet able to explain the observed broadband SED, but also that that a jet model cannot be precluded simply based on the shape of the X-ray spectrum.

\subsection{Comparison with other sources}
\label{sec:disccomp}

Here we compare the \m94 model parameters with approximate values and ranges previously obtained within the \citetalias{MarkoffNowakWilms2005} framework, for BHB states as well as for individual AGN. These values and ranges are summarised in Table \ref{tab:prevval}. 

\begin{table*}\centering
%\scriptsize
\begin{tabular}{lcccccccccc}
\cline{3-11}
							&& 			\mc{5}{c}{AGN}													&& \mc{3}{c}{BHBs}												\\
							&& \4051				&M81*			& \mc{2}{c}{\sgra}			& M87		&& quiescence			& HS				& $\chi$-state 				\\
\cline{1-1}\cline{3-7}\cline{9-11}	
\njet ($10^{-5}$ \ledd)			&&$2\times10^{3}$		&$4$				&0.03		& 0.04		&$0.05-0.2$	&& 1 --146			& 30--8400		& 0.2--2$\times10^5$		\\	
\te ($10^{11}$ K)				&& 0.4				& 1				& 3			& 4			&3--4.5		&& 0.1--0.3			& 0.3--0.7			&0.04 -- 0.09				\\		
\rzero (\rg)						&& 4					& 5				& 4			& 2			&3--10		&& 2--4				& 3--114			& 4--20					\\
$k$							&& 1*				& 1--2			& 1.5			& 0.1			&0.3--1.1		&& 0.044--2			& 1.1--5.4			&29--707					\\
\escat (10$^{-4}$)				&& --					& 5--40			& $10^{2}$	& $10^{-2}$	&$10^{-3}$*	&& 0.01--25			& 1-440			&0.5--1.7 					\\	
$p$							&& 2.7				& 2.4				& 2.5			& 2.6			&2.5--3.0		&& 2.2--3.0			& 2.2-2.9			&1.2--2.5					\\	
\zacc (\rg)						&& 5					& 130--200		& 40			& 23			&8--50		&& 14--25				& 7--526			&7--844 $\times10^{3}$		\\
\hratio						&& 1					& 5--15			& 1.5*		& 1.3			&2--10		&& 1.5*				& 0.4--1.7			& 1.5*					\\
\cline{1-1}\cline{3-7}\cline{9-11}	
$\log$(\mbh/\msol)				&& 6.2				& 7.8				& \mc{2}{c}{6.6}				&9.8			&& $\sim1$ 			& $\sim1$ 		& 1.2						\\
\lx (10$^{-5}$ \ledd)				&& $\sim10^{2}$		& 2--6			& \mc{2}{c}{10$^{-4}$}		&10$^{-2}$	&&$<0.1$			& 0.1--10$^{3}$		& $\sim10^5$				\\
\cline{1-1}\cline{3-7}\cline{9-11}	
Dom. X-ray 					&&\mr{2}{*}{syn} 		& syn/			&\mr{2}{*}{syn}	&\mr{2}{*}{SSC}&\mr{2}{*}{SSC}&& syn/				& mostly			& SCC			 		\\
rad. mech   					&&				 	& SSC			&			&			&			&& SSC				& SSC			& preferred				\\
\cline{1-1}\cline{3-7}\cline{9-11}	
References					&&	a				& b				& \mc{2}{c}{c, d, e}			& f			&& g, h				& i, j, k			& l						\\
\end{tabular}
\caption{Summary of approximate fitted values and parameter ranges for BHBs states and individual AGN from previous \citetalias{MarkoffNowakWilms2005} works. Parameters listed with an asterisk (*) were frozen to the listed value. Quiescent BBHs are \a0620 and \xtej, HS BHBs are \gx, \groj and Cyg X-1, the $\chi$ state BH is \grs. For \sgra we are quoting results from \citet[][in prep.]{Connors+2016}, as these these are the best ones available related to (relatively) high luminosity flares (during which \sgra falls on the FP, see text). Also listed are the dominant X-ray radiation mechanism (dom. X-ray rad. mech.) for each AGN/BH state. For \sgra values for both synchrotron and SSC dominated fit are shown explicitly. References: a) \citet{Maitra+2011}, b) \citet{Markoff+2008}, c) \citet{Connors+2016}, d)\citet{MarkoffBowerFalcke2007}, e) \citet{Maitra+2009b}), f) \citet{Prieto+2016}, g) \citet{Gallo+2007}, h) \citet{Plotkin+2015}, i) \citetalias{MarkoffNowakWilms2005}, j) \citet{MaitraMarkoffBrocksopp2009}), k) \citet{Migliari+2007}, l) \citet{vanOers+2010}.}
\label{tab:prevval}
\end{table*}      

The BHs listed in Table \ref{tab:prevval} vary considerably in luminosity ( in Eddington units) and the normalised power going into the jets (\njet) reflects these luminsities, within about an order of magnitude. This relationship is however not expected to be linear as \njet also needs to provide e.g. the kinetic proton energy; an unknown. The normalised jet powers we obtained for \m94 are very similar to values obtained for M81*, as both M81* and M94* are classifiable as LINERs and they are of similar \lledd luminosity. Interestingly, however, the jet in \4051 is rather more powerful even though \4051 is a Seyfert. As described in \citet{Maitra+2011}, \4051 may actually be in a transitional state between soft and hard (assuming we can think of AGN in terms of BHB spectral states). For \sgra the difference between \njet and \lx/\ledd is rather large. \sgra may generally not even be classified as an AGN \citep[e.g.][]{Goldwurm+1994,MastichiadisOzernoy1994}, due to its extremely low luminosity \citep[e.g.][]{Baganoff+2003}. Only in its flaring mode do we observe a non-thermal X-ray spectrum, that increases the X-ray luminosities to fundamental plane values during the brightest flares only \citep{Neilsen+2013,Connors+2016}. The \lx/\ledd value we quote in Table \ref{tab:prevval} is inferred from these non-thermal flares.

At $\sim10^{11}$ K, the electron temperatures obtained in both our models are fully consistent with AGN of similar BH mass previously studied within the \citetalias{MarkoffNowakWilms2005} framework as well as with the inner flow of other RIAF models for other LLAGN \citep[e.g.][]{Yuan+2002,Yuan+2002_4258,Yuan+2006,Moscibrodzka+2016}. The electron temperatures in the different AGN and BHB states reveal an anti-correlation with the luminosities in Eddington that is qualitatively similar to such correlations found for thermal Comptonisation models \citep[e.g.][]{Miyakawa+2008,Niedzwiecki+2014}. There is an indication of a correlation between \te and black hole mass, with the BHBs displaying the lowest temperatures and, on the other end of the mass scale, M87 showing the highest mass and the highest electron temperatures. This could be understood if pair production acts as a means for cooling the inner flow \citep[e.g.][]{Begelman+1987}, as AGN conditions permit higher electron temperatures before starting pair production. However significant pair production is only expected at high luminosity and high compactness \citep{MalzacBelmontFabian2009}.

The \m94 fits confirm the trend seen before that within the \citetalias{MarkoffNowakWilms2005} framework, SMBHs and quiescent BHs show a compact acceleration region, with nozzle radii of only a few \rg. Similarly, all AGN and quiescent BHBs show equipartition between the magnetic and particle energy densities. The HS BHBs, however, have jets tending towards Poynting flux dominated, with the most luminous BHB \grs displaying the greatest magnetic domination, suggesting a correlation between magnetic domination and normalised jet power for BHBs. \citet{Markoff+2008} find evidence for a similar correlation in the case of M81* but for the entire AGN sample this evidence is lacking. However we need to bear in mind that this AGN sample is still rather small and modelling was not performed homogeneously. In the future we will be reconsidering all these issues in the context of a new semi-analytical relativistic MHD jet model that is under development, based on work in \citet{Polko+2010,Polko+2013,Polko+2014}. In summary these trends could be real or they could be artefacts of having too many free parameters at this stage and since the internal pressures are currently decoupled from the jet dynamics. In future work all these parameters will be set up self-consistently.

Interestingly, a more compact jet base together with a sub-equipartition accretion flow (and slightly higher electron temperatures) were also obtained in the SSC dominated fits of the quiescent BHs \a0620 and \xtej, when compared to their equivalent synchrotron dominated fits \citep[][respectively]{Gallo+2007,Plotkin+2015}. An equivalent duo of modelling interpretations is seen for \sgra, where the absolute differences in these three parameters are remarkably similar to those seen in \m94. Unfortunately, for the other very low luminosity LINER 2, M87, only SSC dominated models were presented \citep{Prieto+2016}. Judging from the M87 spectra, it appears possible to model this source with synchrotron dominated X-rays. Doing so would provide further insight into the self-similarity witnessed above for the other lowest accretion rate sources. In the higher luminosity M81* similar statistically equivalent SSC/synchrotron model interpretations are not obtained, but there a fundamentally different model is assumed in the SSC dominated fits, where power-law leptons are injected straight into the nozzle. This change in assumptions precludes a direct comparison between the SSC models of \m94 and M81*. In any case for the lowest luminosity sources it appears our current modelling can not unequivocally distinguish the dominant radiative mechanism in the X-rays. Some works suggest an SSC origin for the lowest Eddington accretion rate sources \citep{Markoff+2001,Plotkin+2015,Prieto+2016}, and a synchrotron origin at higher accretion rates ($10^{-4}\lesssim$\lx/\ledd$\lesssim10^{-3}$; \citealt{Russell+2010}). In line with these results a \xtej transition from the HS to quiescence suggested the preferred model should change from synchrotron to SSC dominated as the Eddington rate decreased \citep{MaitraMarkoffBrocksopp2009,Plotkin+2015}. More such evidence comes from the fact that for the highest accretion rate AGN studied with \citetalias{MarkoffNowakWilms2005}, the Seyfert \4051, only synchrotron dominated fits were deemed feasible \citep{MaitraMarkoffBrocksopp2009}. However this discussion is far from final, particularly considering the Eddington-rate BHB \grs prefers SSC \citep{vanOers+2010}.  

In terms of the particle distribution index $p$, we see general agreement among all sources, with $2<p<3$. In our current modelling we have included a synchrotron cooling break and adequate fits were found at $p<2.5$, consistent with \citet[][]{HeavensDrury1988}. Older values of $p>2.5$ from earlier applications of the model were always implicitly assumed to reflect a cooling dominated regime. Therefore it seems likely that if \4051 and M87 (and similarly, high-$p$ BHBs) were re-explored with a cooling break inclusive model, reduced values for $p$ would be obtained. However if a break steepens the model spectrum at UV-to-X-ray energies significantly, a different location of the acceleration \zacc is implied. Hence a direct comparison of \m94 with previous works, of the optically thick to optically thin turnover (and possible dependencies of it on other parameters), is impossible. For BHBs a correlation between \zacc and \njet (or, equivalently, and perhaps more importantly, an anti-correlation between mass accretion rate and the break frequency) is seen \citep{Russell+2013}. For the LLAGN (excluding the ``intermediate state" \4051) we see a similar evolution, however due to the small number of available AGN the evidence is not so clear: The location of the acceleration obtained for \m94 are very similar to those found for M81*,  especially considering that jets cover up to $10^7$ \rg and hence these distances are preferably considered in log-space. Indeed, the fact that all the values in Table \ref{tab:prevval} fall within $\log$(\zacc) $\sim1$--4 for all black holes of any size is already interesting in itself. Re-modelling of the AGN with a cooling break inclusive model would however be necessary before drawing firm conclusions on any possible relations with \zacc. Determining the distance to the acceleration region self-consistently in a jet from theory has been the subject of the work mentioned above by \citet{Polko+2010,Polko+2013,Polko+2014} and is currently under further development (Ceccobello et al. in prep.).

To summarise, the above discussion suggests that the (pure) LINERs previously studied with \citetalias{MarkoffNowakWilms2005} may be closer in nature to quiescent BHBs than to HS BHBs, perhaps unsurprisingly, as these LLAGN display the lowest accretion rates of the AGN class: Compared to their high ionisation counterparts, for a similar galaxy type and stellar mass (that is strongly linked to the central BH mass; \citealt{FerrareseMerritt2000,Gebhardt+2000}), LINERs typically display nuclear luminosities 1 to 5 orders of magnitude lower \citep{Netzerbook}. However we must bear in mind that the LLAGN discussed here are very similar in Eddington accretion rate to quiescent BHBs, which is a more important factor than the absolute luminosity. In fact, \citet{Markoff+2015,Connors+2016} show that when using such mass-invariant accretion rates, \citetalias{MarkoffNowakWilms2005} works well across the entire BH mass scale. 

However, as indicated above, from our results there appear to be emerging subtle differences between the AGN and BHB classes, in terms of the level of equipartition and normalised jet power and the dominant radiation mechanism in the X-rays at higher accretion rates. In order to investigate these matters it will certainly help to explore more AGN at the higher end of the accretion rate scale. Only a single NLS1 has been modelled in the context of the MNW05 model up to this point and it would be insightful to see if other radio loud Seyferts would permit SSC or also require synchrotron domination. Furthermore, the AGN and BHB classes show a number of major observational differences: In BHBs, after a hard-to-soft transition (where optically thin relativistic ejecta are launched; e.g. \citealt{Fender+1999,Fender+2004}), no steady jets are observed, only radiatively-driven winds \citep[e.g.][]{NeilsenLee2009,Ponti+2012}. For AGN this is not the case: both slow, steady jets (like those found in the hard state of XRBs) and high-power, relativistic jets (like those observed in the hard-to-soft transition in BHBs) can be observed in the radiatively efficient, soft-state-equivalent AGN, sometimes coexisting with winds as well (e.g. \citealt{Nesvadba+2008,Mullaney+2013,Collet+2016}). Although there are various theories, attributing various mechanisms for this observed difference between BHBs and AGN (e.g. \citealt{Tchekhovskoy+2011,Hardcastle+2007,SikoraBegelman2013,NemmenTchekhovskoy2015,Tadhunter2016}), we are far from consensus about the definitive mechanism. Applying the model from MNW05 to other radio-loud Seyferts could help to understand what is causing this difference. Lastly, multi-wavelength quasar SEDs \citep[e.g.][]{Elvis+1994} are necessary to fully explore the higher end of the SMBH accretion range.

Indeed, modelling high accretion rates is not straight forward and other problems will have to be taken into account: First of all we need to account for pair-production in the model. Secondly, at high luminosity the accretion flow is thought to turn efficient, while previously the main objective of \citetalias{MarkoffNowakWilms2005} has been to study the low accretion rate, inefficient accretion flows that fall onto the fundamental plane, as these are unambiguously associated with jets. Although at the highest accretion rates the flow may turn inefficient again, these may be regimes for which an \citetalias{MarkoffNowakWilms2005} jet is not appropriate. But with these caveats, studying higher accretion rate sources and especially ``intermediate states" (like \grs and \4051) should be feasible. Such modelling would allow further investigation of the fundamental plane, particularly whether the high accretion rate sources fall onto the appropriate loci extending the efficient track to higher luminosities, or whether they perhaps draw an altogether different radio/X-ray correlation track \citep[e.g.][]{Dong+2014,XieYuan2016}, possibly providing complementary information on the underlying driving physics and radiation mechanisms.

%%%%%%%%%%%%%%%%%%%%%%%%%%%%%%%%%%%%%%%%%%%%
%%%%%%%%%%%%%%%%%%%%%%%%%%%%%%%%%%%%%%%%%%%%%%%%%
%CONCL%%%%%%%%%%%%%%%%%%%%%%%%%%%%%%%%%%%%%%
%%%%%%%%%%%%%%%%%%%%%%%%%%%%%%%%%%%%%%%%%%%%%%%%

\section{Conclusions}
\label{sec:nconcl}

We presented a broadband SED of the nucleus of the closest obscured LINER 2, M94, composed of predominantly sub-arcsec resolution data, and modelled the SED with an outflow-dominated model. The results have shown that even though the X-ray spectrum is relatively steep, a jet/diffusive shock acceleration can adequately explain the data. If synchrotron dominated X-rays are assumed, a cooling break is found in that energy regime. With respect to electron temperature, compactness and equipartition in the inner flow our results are completely consistent with previous finding for LLAGN and quiescent BHBs, but unfortunately we also recover and were unable to break commonly found degeneracies, in terms of the dominant radiation mechanisms. 

Most importantly however, work done here increases the number of sources that have been explored with jet-dominated SEDs, and the growing number allows for increasingly insightful comparisons between different classes of BHs that differ many magnitudes in mass. It is worth noting, however, that especially the high accretion rate regime remains to be investigated in more depth. In this regime different problems such as pair-production, as well as presumably an altogether different type of accretion will require further expansion of the frameworks used. Work in that regard is however well-underway.

Furthermore, as a myriad of new telescope arrays are currently being developed and built, the quality of available SEDs is certain to improve by orders of magnitude, reaching new levels of detail, complexity and signal-to-noise. Moreover the impending \emph{transient era} \citep{Metzger+2015} is sure to deliver a wealth of new information on jetted sources: As wide field observatories like the Square Kilometer Array, the Cherenkov Telescope Array and the Large Synoptic Survey Telescope will provide large-footprint surveys with a high frequency of repeated observations, yielding accurate time-sampling for BHBs as they go through their outburst and quiescent cycles. Even in the near future improvements can be expected for one of the arrays used in this work: \emerlin should soon meet its design specifications, sporting a broad (2 GHz) bandwidth at C-band (5 GHz). This will allow us not only to study in detail the short-term radio variability by verifying whether the \m94 flat/inverted spectrum is truly as uniform as we see in this work, or whether - because of e.g. absorption - the uniformity breaks down when resolving the radio frequencies into smaller bandwidth pieces (in a way similar to what we observed here in our currently available L-band data), but also to investigate the long-term variability, comparing near-future flux-levels in \m94 to those of the oldest data in our data set, namely the data around 5 GHz that are currently over 3 decades old.

Clearly the study presented here would benefit from the addition of (sub)millimetre data. Data from the Atacama Large Millimeter/submillimeter Array, for example, would be extremely beneficial in breaking degeneracies in the modelling, as this would help us in ascertaining the exact energy and shape of one of the most revealing features in our model spectrum; the ``sub-mm bump". Considering the importance of this feature in jet physics and the sensitivity of this feature on other jet parameters, having this data available would surely allow us to preclude a number of interpretations as viable solutions. In addition, sub-mm and mm wavelengths have the added benefit that they offer the highest angular resolution in this band. In particular, most LLAGN turn optically thin and are more transparent at shorter wavelength, starting in the (sub)mm band. VLBI at these higher frequencies would allow observation of the core regions of the accretion zone that are hidden from view at lower radio frequencies, ultimately even allowing direct imaging of the central engine of AGN. Although it will not be able to resolve the core in M94, the submm observations the Event Horizon Telescope \citep{Honma+2016} will make when it comes online in a few years are sure to provide a wealth of new insight into the nature of accretion flows, jets and AGN, as it should for the first time be able to glimpse what exactly is happening near the edge of a BH.

In the meantime any authoritative model would need to tackle issues of degeneracy and deal with these issues with a self consistent approach. The ground work for this has been laid in \citet{Polko+2010,Polko+2013,Polko+2014}. However until implementations of these papers are complete (Ceccobello et al. in prep.), individual source studies with currently available frameworks -- as done here -- will help discern trends via comparative studies and pinpoint the relevant parameter space for this next generation of all-encompassing fully self-consistent magnetohydrodynamic models.

\section{Acknowledgements}

We thank Pierre-Emmanuel Belles and Rob Beswick for assistance reducing the \emerlin radio images.
We thank the referee and Thomas Russell for useful comments.
The NVAS images used were produced as part of the NRAO VLA Archive Survey, (c) AUI/NRAO. 
The National Radio Astronomy Observatory is a facility of the National Science Foundation operated under cooperative agreement by Associated Universities, Inc.

\bibliographystyle{mnras}
\bibliography{/Users/pieter/Dropbox/science/citations/citationthes}

\newcommand{\noop}[1]{}
\begin{thebibliography}{}
\makeatletter
\relax
\def\mn@urlcharsother{\let\do\@makeother \do\$\do\&\do\#\do\^\do\_\do\%\do\~}
\def\mn@doi{\begingroup\mn@urlcharsother \@ifnextchar [ {\mn@doi@}
  {\mn@doi@[]}}
\def\mn@doi@[#1]#2{\def\@tempa{#1}\ifx\@tempa\@empty \href
  {http://dx.doi.org/#2} {doi:#2}\else \href {http://dx.doi.org/#2} {#1}\fi
  \endgroup}
\def\mn@eprint#1#2{\mn@eprint@#1:#2::\@nil}
\def\mn@eprint@arXiv#1{\href {http://arxiv.org/abs/#1} {{\tt arXiv:#1}}}
\def\mn@eprint@dblp#1{\href {http://dblp.uni-trier.de/rec/bibtex/#1.xml}
  {dblp:#1}}
\def\mn@eprint@#1:#2:#3:#4\@nil{\def\@tempa {#1}\def\@tempb {#2}\def\@tempc
  {#3}\ifx \@tempc \@empty \let \@tempc \@tempb \let \@tempb \@tempa \fi \ifx
  \@tempb \@empty \def\@tempb {arXiv}\fi \@ifundefined
  {mn@eprint@\@tempb}{\@tempb:\@tempc}{\expandafter \expandafter \csname
  mn@eprint@\@tempb\endcsname \expandafter{\@tempc}}}

\bibitem[\protect\citeauthoryear{{Arnaud}}{{Arnaud}}{1996}]{Arnaud1996}
{Arnaud} K.~A.,  1996, in {Jacoby} G.~H.,  {Barnes} J.,  eds,  Astronomical
  Society of the Pacific Conference Series Vol. 101, Astronomical Data Analysis
  Software and Systems V. pp 17--+

\bibitem[\protect\citeauthoryear{{Asmus}, {H{\"o}nig}, {Gandhi}, {Smette}  \&
  {Duschl}}{{Asmus} et~al.}{2014}]{Asmus+2014}
{Asmus} D.,  {H{\"o}nig} S.~F.,  {Gandhi} P.,  {Smette} A.,   {Duschl} W.~J.,
  2014, \mn@doi [\mnras] {10.1093/mnras/stu041}, \href
  {http://adsabs.harvard.edu/abs/2014MNRAS.439.1648A} {439, 1648}

\bibitem[\protect\citeauthoryear{{Baganoff} et~al.,}{{Baganoff}
  et~al.}{2003}]{Baganoff+2003}
{Baganoff} F.~K.,  et~al., 2003, \mn@doi [\apj] {10.1086/375145}, \href
  {http://adsabs.harvard.edu/abs/2003ApJ...591..891B} {591, 891}

\bibitem[\protect\citeauthoryear{{Baldwin}, {Phillips}  \&
  {Terlevich}}{{Baldwin} et~al.}{1981}]{BaldwinPhillipsTerlevich1981}
{Baldwin} J.~A.,  {Phillips} M.~M.,   {Terlevich} R.,  1981, \mn@doi [\pasp]
  {10.1086/130766}, \href {http://adsabs.harvard.edu/abs/1981PASP...93....5B}
  {93, 5}

\bibitem[\protect\citeauthoryear{{Begelman}, {Sikora}  \& {Rees}}{{Begelman}
  et~al.}{1987}]{Begelman+1987}
{Begelman} M.~C.,  {Sikora} M.,   {Rees} M.~J.,  1987, \mn@doi [\apj]
  {10.1086/165007}, \href {http://adsabs.harvard.edu/abs/1987ApJ...313..689B}
  {313, 689}

\bibitem[\protect\citeauthoryear{{Belloni}}{{Belloni}}{2010}]{Belloni2010}
{Belloni} T.~M.,  2010, in {Belloni} T.,  ed.,  Lecture Notes in Physics,
  Berlin Springer Verlag Vol. 794, Lecture Notes in Physics, Berlin Springer
  Verlag. p.~53 (\mn@eprint {arXiv} {0909.2474}),
  \mn@doi{10.1007/978-3-540-76937-8_3}

\bibitem[\protect\citeauthoryear{{Beloborodov}}{{Beloborodov}}{1999}]{Beloborodov1999}
{Beloborodov} A.~M.,  1999, \mn@doi [\apjl] {10.1086/311810}, \href
  {http://adsabs.harvard.edu/abs/1999ApJ...510L.123B} {510, L123}

\bibitem[\protect\citeauthoryear{{Biretta}, {Stern}  \& {Harris}}{{Biretta}
  et~al.}{1991}]{Biretta+1991}
{Biretta} J.~A.,  {Stern} C.~P.,   {Harris} D.~E.,  1991, \mn@doi [\aj]
  {10.1086/115793}, \href {http://adsabs.harvard.edu/abs/1991AJ....101.1632B}
  {101, 1632}

\bibitem[\protect\citeauthoryear{{B{\^i}rzan}, {Rafferty}, {McNamara}, {Wise}
  \& {Nulsen}}{{B{\^i}rzan} et~al.}{2004}]{Birzan+2004}
{B{\^i}rzan} L.,  {Rafferty} D.~A.,  {McNamara} B.~R.,  {Wise} M.~W.,
  {Nulsen} P.~E.~J.,  2004, \mn@doi [\apj] {10.1086/383519}, \href
  {http://adsabs.harvard.edu/abs/2004ApJ...607..800B} {607, 800}

\bibitem[\protect\citeauthoryear{{Blandford} \& {Konigl}}{{Blandford} \&
  {Konigl}}{1979}]{BlandfordKonigl1979}
{Blandford} R.~D.,  {Konigl} A.,  1979, \mn@doi [\apj] {10.1086/157262}, \href
  {http://adsabs.harvard.edu/abs/1979ApJ...232...34B} {232, 34}

\bibitem[\protect\citeauthoryear{{Bruzual} \& {Charlot}}{{Bruzual} \&
  {Charlot}}{2003}]{BruzualCharlot2003}
{Bruzual} G.,  {Charlot} S.,  2003, \mn@doi [\mnras]
  {10.1046/j.1365-8711.2003.06897.x}, \href
  {http://adsabs.harvard.edu/abs/2003MNRAS.344.1000B} {344, 1000}

\bibitem[\protect\citeauthoryear{{Buta}, {Corwin}  \& {Odewahn}}{{Buta}
  et~al.}{2007}]{Buta+2007}
{Buta} R.~J.,  {Corwin} H.~G.,   {Odewahn} S.~C.,  2007, {The de Vaucouleurs
  Altlas of Galaxies}.
Cambridge University Press

\bibitem[\protect\citeauthoryear{{Calzetti}, {Kinney}  \&
  {Storchi-Bergmann}}{{Calzetti} et~al.}{1994}]{Calzetti1994}
{Calzetti} D.,  {Kinney} A.~L.,   {Storchi-Bergmann} T.,  1994, \mn@doi [\apj]
  {10.1086/174346}, \href {http://adsabs.harvard.edu/abs/1994ApJ...429..582C}
  {429, 582}

\bibitem[\protect\citeauthoryear{{Carilli}, {Perley}, {Dreher}  \&
  {Leahy}}{{Carilli} et~al.}{1991}]{Carilli+1991}
{Carilli} C.~L.,  {Perley} R.~A.,  {Dreher} J.~W.,   {Leahy} J.~P.,  1991,
  \mn@doi [\apj] {10.1086/170813}, \href
  {http://adsabs.harvard.edu/abs/1991ApJ...383..554C} {383, 554}

\bibitem[\protect\citeauthoryear{{Ceccobello et al.}}{{Ceccobello et
  al.}}{prep}]{Ceccobello+2017}
{Ceccobello et al.} C.,  "in prep.", \mnras

\bibitem[\protect\citeauthoryear{{Clemens}, {Vega}, {Bressan}, {Granato},
  {Silva}  \& {Panuzzo}}{{Clemens} et~al.}{2008}]{Clemens+2008}
{Clemens} M.~S.,  {Vega} O.,  {Bressan} A.,  {Granato} G.~L.,  {Silva} L.,
  {Panuzzo} P.,  2008, \mn@doi [\aap] {10.1051/0004-6361:20077224}, \href
  {http://adsabs.harvard.edu/abs/2008A%26A...477...95C} {477, 95}

\bibitem[\protect\citeauthoryear{{Collet} et~al.,}{{Collet}
  et~al.}{2016}]{Collet+2016}
{Collet} C.,  et~al., 2016, \mn@doi [\aap] {10.1051/0004-6361/201526872}, \href
  {http://adsabs.harvard.edu/abs/2016A%26A...586A.152C} {586, A152}

\bibitem[\protect\citeauthoryear{{Condon}}{{Condon}}{1992}]{Condon1992}
{Condon} J.~J.,  1992, \mn@doi [\araa] {10.1146/annurev.aa.30.090192.003043},
  \href {http://adsabs.harvard.edu/abs/1992ARA%26A..30..575C} {30, 575}

\bibitem[\protect\citeauthoryear{{Connors} et~al.,}{{Connors}
  et~al.}{2016}]{Connors+2016}
{Connors} R.~M.~T.,  et~al., 2016, preprint, \href
  {http://adsabs.harvard.edu/abs/2016arXiv161200953C} {} (\mn@eprint {arXiv}
  {1612.00953})

\bibitem[\protect\citeauthoryear{{Constantin} \& {Seth}}{{Constantin} \&
  {Seth}}{2012}]{Constantin+2012}
{Constantin} A.,  {Seth} A.~C.,  2012, \mn@doi [Advances in Astronomy]
  {10.1155/2012/178060}, \href
  {http://adsabs.harvard.edu/abs/2012AdAst2012E..13C} {2012, 13}

\bibitem[\protect\citeauthoryear{{Doane}, {Sanders}, {Wilcots}  \&
  {Juda}}{{Doane} et~al.}{2004}]{Doane+2004}
{Doane} N.~E.,  {Sanders} W.~T.,  {Wilcots} E.~M.,   {Juda} M.,  2004, \mn@doi
  [\aj] {10.1086/425627}, \href
  {http://adsabs.harvard.edu/abs/2004AJ....128.2712D} {128, 2712}

\bibitem[\protect\citeauthoryear{{Dong}, {Wu}  \& {Cao}}{{Dong}
  et~al.}{2014}]{Dong+2014}
{Dong} A.-J.,  {Wu} Q.,   {Cao} X.-F.,  2014, \mn@doi [\apjl]
  {10.1088/2041-8205/787/2/L20}, \href
  {http://adsabs.harvard.edu/abs/2014ApJ...787L..20D} {787, L20}

\bibitem[\protect\citeauthoryear{{Donovan Meyer} et~al.,}{{Donovan Meyer}
  et~al.}{2013}]{DonovanMeyer+2013}
{Donovan Meyer} J.,  et~al., 2013, \mn@doi [\apj]
  {10.1088/0004-637X/772/2/107}, \href
  {http://adsabs.harvard.edu/abs/2013ApJ...772..107D} {772, 107}

\bibitem[\protect\citeauthoryear{{Dopita} et~al.,}{{Dopita}
  et~al.}{2005}]{Dopita+2005}
{Dopita} M.~A.,  et~al., 2005, \mn@doi [\apj] {10.1086/423948}, \href
  {http://adsabs.harvard.edu/abs/2005ApJ...619..755D} {619, 755}

\bibitem[\protect\citeauthoryear{{Elvis} et~al.,}{{Elvis}
  et~al.}{1994}]{Elvis+1994}
{Elvis} M.,  et~al., 1994, \mn@doi [\apjs] {10.1086/192093}, \href
  {http://adsabs.harvard.edu/abs/1994ApJS...95....1E} {95, 1}

\bibitem[\protect\citeauthoryear{{Eracleous}, {Shields}, {Chartas}  \&
  {Moran}}{{Eracleous} et~al.}{2002}]{Eracleous+2002}
{Eracleous} M.,  {Shields} J.~C.,  {Chartas} G.,   {Moran} E.~C.,  2002,
  \mn@doi [\apj] {10.1086/324394}, \href
  {http://adsabs.harvard.edu/abs/2002ApJ...565..108E} {565, 108}

\bibitem[\protect\citeauthoryear{{Eracleous}, {Hwang}  \& {Flohic}}{{Eracleous}
  et~al.}{2010}]{Eracleous+2010}
{Eracleous} M.,  {Hwang} J.~A.,   {Flohic} H.~M.~L.~G.,  2010, \mn@doi [\apjs]
  {10.1088/0067-0049/187/1/135}, \href
  {http://adsabs.harvard.edu/abs/2010ApJS..187..135E} {187, 135}

\bibitem[\protect\citeauthoryear{{Erwin}}{{Erwin}}{2004}]{Erwin2004}
{Erwin} P.,  2004, \mn@doi [\aap] {10.1051/0004-6361:20034408}, \href
  {http://adsabs.harvard.edu/abs/2004A%26A...415..941E} {415, 941}

\bibitem[\protect\citeauthoryear{{Esin}, {McClintock}  \& {Narayan}}{{Esin}
  et~al.}{1997}]{Esin+1997}
{Esin} A.~A.,  {McClintock} J.~E.,   {Narayan} R.,  1997, \mn@doi [\apj]
  {10.1086/304829}, \href {http://adsabs.harvard.edu/abs/1997ApJ...489..865E}
  {489, 865}

\bibitem[\protect\citeauthoryear{{Falcke}}{{Falcke}}{1996}]{Falcke1996}
{Falcke} H.,  1996, \mn@doi [\apjl] {10.1086/310085}, \href
  {http://adsabs.harvard.edu/abs/1996ApJ...464L..67F} {464, L67+}

\bibitem[\protect\citeauthoryear{{Falcke}, {K{\"o}rding}  \&
  {Markoff}}{{Falcke} et~al.}{2004}]{Falcke+2004}
{Falcke} H.,  {K{\"o}rding} E.,   {Markoff} S.,  2004, \mn@doi [\aap]
  {10.1051/0004-6361:20031683}, \href
  {http://adsabs.harvard.edu/abs/2004A%26A...414..895F} {414, 895}

\bibitem[\protect\citeauthoryear{{Fender}, {Garrington}, {McKay}, {Muxlow},
  {Pooley}, {Spencer}, {Stirling}  \& {Waltman}}{{Fender}
  et~al.}{1999}]{Fender+1999}
{Fender} R.~P.,  {Garrington} S.~T.,  {McKay} D.~J.,  {Muxlow} T.~W.~B.,
  {Pooley} G.~G.,  {Spencer} R.~E.,  {Stirling} A.~M.,   {Waltman} E.~B.,
  1999, \mnras, \href {http://adsabs.harvard.edu/abs/Fender+1999} {304, 865}

\bibitem[\protect\citeauthoryear{{Fender}, {Belloni}  \& {Gallo}}{{Fender}
  et~al.}{2004}]{Fender+2004}
{Fender} R.~P.,  {Belloni} T.~M.,   {Gallo} E.,  2004, \mn@doi [\mnras]
  {10.1111/j.1365-2966.2004.08384.x}, \href
  {http://adsabs.harvard.edu/abs/2004MNRAS.355.1105F} {355, 1105}

\bibitem[\protect\citeauthoryear{{Fern{\'a}ndez-Ontiveros}, {Prieto},
  {Acosta-Pulido}  \& {Montes}}{{Fern{\'a}ndez-Ontiveros}
  et~al.}{2012}]{Fernandez-Ontiveros+2012}
{Fern{\'a}ndez-Ontiveros} J.~A.,  {Prieto} M.~A.,  {Acosta-Pulido} J.~A.,
  {Montes} M.,  2012, \mn@doi [Journal of Physics Conference Series]
  {10.1088/1742-6596/372/1/012006}, \href
  {http://adsabs.harvard.edu/abs/2012JPhCS.372a2006F} {372, 012006}

\bibitem[\protect\citeauthoryear{{Ferrarese} \& {Merritt}}{{Ferrarese} \&
  {Merritt}}{2000}]{FerrareseMerritt2000}
{Ferrarese} L.,  {Merritt} D.,  2000, \mn@doi [\apjl] {10.1086/312838}, \href
  {http://adsabs.harvard.edu/abs/2000ApJ...539L...9F} {539, L9}

\bibitem[\protect\citeauthoryear{{Fruscione} et~al.,}{{Fruscione}
  et~al.}{2006}]{Fruscione+2006}
{Fruscione} A.,  et~al., 2006, in Society of Photo-Optical Instrumentation
  Engineers (SPIE) Conference Series. p.~1, \mn@doi{10.1117/12.671760}

\bibitem[\protect\citeauthoryear{{Gallo}, {Migliari}, {Markoff}, {Tomsick},
  {Bailyn}, {Berta}, {Fender}  \& {Miller-Jones}}{{Gallo}
  et~al.}{2007}]{Gallo+2007}
{Gallo} E.,  {Migliari} S.,  {Markoff} S.,  {Tomsick} J.~A.,  {Bailyn} C.~D.,
  {Berta} S.,  {Fender} R.,   {Miller-Jones} J.~C.~A.,  2007, \mn@doi [\apj]
  {10.1086/521524}, \href {http://adsabs.harvard.edu/abs/2007ApJ...670..600G}
  {670, 600}

\bibitem[\protect\citeauthoryear{{Garrington} et~al.,}{{Garrington}
  et~al.}{2004}]{Garrington+2004}
{Garrington} S.~T.,  et~al., 2004, in {Oschmann} Jr. J.~M.,  ed.,  \procspie
  Vol. 5489, Ground-based Telescopes. pp 332--343, \mn@doi{10.1117/12.553235}

\bibitem[\protect\citeauthoryear{{Gebhardt} et~al.,}{{Gebhardt}
  et~al.}{2000}]{Gebhardt+2000}
{Gebhardt} K.,  et~al., 2000, \mn@doi [\apjl] {10.1086/312840}, \href
  {http://adsabs.harvard.edu/abs/2000ApJ...539L..13G} {539, L13}

\bibitem[\protect\citeauthoryear{{Gilfanov}}{{Gilfanov}}{2010}]{Gilfanov2010}
{Gilfanov} M.,  2010, in {Belloni} T.,  ed.,  Lecture Notes in Physics, Berlin
  Springer Verlag Vol. 794, Lecture Notes in Physics, Berlin Springer Verlag.
  p.~17 (\mn@eprint {arXiv} {0909.2567}), \mn@doi{10.1007/978-3-540-76937-8_2}

\bibitem[\protect\citeauthoryear{{Gillessen}, {Eisenhauer}, {Fritz}, {Bartko},
  {Dodds-Eden}, {Pfuhl}, {Ott}  \& {Genzel}}{{Gillessen}
  et~al.}{2009}]{Gillessen+2009}
{Gillessen} S.,  {Eisenhauer} F.,  {Fritz} T.~K.,  {Bartko} H.,  {Dodds-Eden}
  K.,  {Pfuhl} O.,  {Ott} T.,   {Genzel} R.,  2009, \mn@doi [\apjl]
  {10.1088/0004-637X/707/2/L114}, \href
  {http://adsabs.harvard.edu/abs/2009ApJ...707L.114G} {707, L114}

\bibitem[\protect\citeauthoryear{{Goldwurm} et~al.,}{{Goldwurm}
  et~al.}{1994}]{Goldwurm+1994}
{Goldwurm} A.,  et~al., 1994, \mn@doi [\nat] {10.1038/371589a0}, \href
  {http://adsabs.harvard.edu/abs/1994Natur.371..589G} {371, 589}

\bibitem[\protect\citeauthoryear{{Greisen}}{{Greisen}}{2003}]{Greisen2003}
{Greisen} E.~W.,  2003, \mn@doi [Information Handling in Astronomy - Historical
  Vistas] {10.1007/0-306-48080-8_7}, \href
  {http://adsabs.harvard.edu/abs/2003ASSL..285..109G} {285, 109}

\bibitem[\protect\citeauthoryear{{Hardcastle}, {Evans}  \&
  {Croston}}{{Hardcastle} et~al.}{2007}]{Hardcastle+2007}
{Hardcastle} M.~J.,  {Evans} D.~A.,   {Croston} J.~H.,  2007, \mn@doi [\mnras]
  {10.1111/j.1365-2966.2007.11572.x}, \href
  {http://adsabs.harvard.edu/abs/2007MNRAS.376.1849H} {376, 1849}

\bibitem[\protect\citeauthoryear{{Heavens} \& {Drury}}{{Heavens} \&
  {Drury}}{1988}]{HeavensDrury1988}
{Heavens} A.~F.,  {Drury} L.~O.,  1988, \mnras, \href
  {http://adsabs.harvard.edu/abs/1988MNRAS.235..997H} {235, 997}

\bibitem[\protect\citeauthoryear{{Ho}}{{Ho}}{1999}]{Ho1999}
{Ho} L.~C.,  1999, \mn@doi [\apj] {10.1086/307137}, \href
  {http://adsabs.harvard.edu/abs/1999ApJ...516..672H} {516, 672}

\bibitem[\protect\citeauthoryear{{Ho} \& {Peng}}{{Ho} \&
  {Peng}}{2001}]{HoPeng2001}
{Ho} L.~C.,  {Peng} C.~Y.,  2001, \mn@doi [\apj] {10.1086/321524}, \href
  {http://adsabs.harvard.edu/abs/2001ApJ...555..650H} {555, 650}

\bibitem[\protect\citeauthoryear{{Ho}, {Filippenko}  \& {Sargent}}{{Ho}
  et~al.}{1997}]{Ho+1997}
{Ho} L.~C.,  {Filippenko} A.~V.,   {Sargent} W.~L.~W.,  1997, \mn@doi [\apjs]
  {10.1086/313041}, \href {http://adsabs.harvard.edu/abs/1997ApJS..112..315H}
  {112, 315}

\bibitem[\protect\citeauthoryear{{Honma}, {Akiyama}, {Tazaki}, {Kuramochi},
  {Ikeda}, {Hada}  \& {Uemura}}{{Honma} et~al.}{2016}]{Honma+2016}
{Honma} M.,  {Akiyama} K.,  {Tazaki} F.,  {Kuramochi} K.,  {Ikeda} S.,  {Hada}
  K.,   {Uemura} M.,  2016, \mn@doi [Journal of Physics Conference Series]
  {10.1088/1742-6596/699/1/012006}, \href
  {http://adsabs.harvard.edu/abs/2016JPhCS.699a2006H} {699, 012006}

\bibitem[\protect\citeauthoryear{{Houck} \& {Denicola}}{{Houck} \&
  {Denicola}}{2000}]{HouckDenicola2000}
{Houck} J.~C.,  {Denicola} L.~A.,  2000, in {Manset} N.,  {Veillet} C.,
  {Crabtree} D.,  eds,  Astronomical Society of the Pacific Conference Series
  Vol. 216, Astronomical Data Analysis Software and Systems IX. pp 591--+

\bibitem[\protect\citeauthoryear{{Hubble}}{{Hubble}}{1926}]{Hubble1926}
{Hubble} E.~P.,  1926, \mn@doi [\apj] {10.1086/143018}, \href
  {http://adsabs.harvard.edu/abs/1926ApJ....64..321H} {64, 321}

\bibitem[\protect\citeauthoryear{{Jensen}, {Tonry}, {Barris}, {Thompson},
  {Liu}, {Rieke}, {Ajhar}  \& {Blakeslee}}{{Jensen} et~al.}{2003}]{Jensen+2003}
{Jensen} J.~B.,  {Tonry} J.~L.,  {Barris} B.~J.,  {Thompson} R.~I.,  {Liu}
  M.~C.,  {Rieke} M.~J.,  {Ajhar} E.~A.,   {Blakeslee} J.~P.,  2003, \mn@doi
  [\apj] {10.1086/345430}, \href
  {http://adsabs.harvard.edu/abs/2003ApJ...583..712J} {583, 712}

\bibitem[\protect\citeauthoryear{{Jokipii}}{{Jokipii}}{1987}]{Jokipii1987}
{Jokipii} J.~R.,  1987, \mn@doi [\apj] {10.1086/165022}, \href
  {http://adsabs.harvard.edu/abs/1987ApJ...313..842J} {313, 842}

\bibitem[\protect\citeauthoryear{{Kardashev}}{{Kardashev}}{1962}]{Kardashev1962}
{Kardashev} N.~S.,  1962, Soviet Astronomy, \href
  {http://adsabs.harvard.edu/abs/1962SvA.....6..317K} {6, 317}

\bibitem[\protect\citeauthoryear{{Kauffmann} et~al.,}{{Kauffmann}
  et~al.}{2003}]{Kauffmann+2003}
{Kauffmann} G.,  et~al., 2003, \mn@doi [\mnras]
  {10.1111/j.1365-2966.2003.07154.x}, \href
  {http://adsabs.harvard.edu/abs/2003MNRAS.346.1055K} {346, 1055}

\bibitem[\protect\citeauthoryear{{Keel} et~al.,}{{Keel}
  et~al.}{2012}]{Keel+2012}
{Keel} W.~C.,  et~al., 2012, \mn@doi [\mnras]
  {10.1111/j.1365-2966.2011.20101.x}, \href
  {http://adsabs.harvard.edu/abs/2012MNRAS.420..878K} {420, 878}

\bibitem[\protect\citeauthoryear{{Keel} et~al.,}{{Keel}
  et~al.}{2015}]{Keel+2015}
{Keel} W.~C.,  et~al., 2015, \mn@doi [\aj] {10.1088/0004-6256/149/5/155}, \href
  {http://adsabs.harvard.edu/abs/2015AJ....149..155K} {149, 155}

\bibitem[\protect\citeauthoryear{{Kewley}, {Groves}, {Kauffmann}  \&
  {Heckman}}{{Kewley} et~al.}{2006}]{Kewley+2006}
{Kewley} L.~J.,  {Groves} B.,  {Kauffmann} G.,   {Heckman} T.,  2006, \mn@doi
  [\mnras] {10.1111/j.1365-2966.2006.10859.x}, \href
  {http://adsabs.harvard.edu/abs/2006MNRAS.372..961K} {372, 961}

\bibitem[\protect\citeauthoryear{{Kinney}, {Schmitt}, {Clarke}, {Pringle},
  {Ulvestad}  \& {Antonucci}}{{Kinney} et~al.}{2000}]{Kinney+2000}
{Kinney} A.~L.,  {Schmitt} H.~R.,  {Clarke} C.~J.,  {Pringle} J.~E.,
  {Ulvestad} J.~S.,   {Antonucci} R.~R.~J.,  2000, \mn@doi [\apj]
  {10.1086/309016}, \href {http://adsabs.harvard.edu/abs/2000ApJ...537..152K}
  {537, 152}

\bibitem[\protect\citeauthoryear{{Koay}, {Vestergaard}, {Bignall}, {Reynolds}
  \& {Peterson}}{{Koay} et~al.}{2016}]{Koay+2016}
{Koay} J.~Y.,  {Vestergaard} M.,  {Bignall} H.~E.,  {Reynolds} C.,   {Peterson}
  B.~M.,  2016, preprint, \href
  {http://adsabs.harvard.edu/abs/2016arXiv160207289K} {} (\mn@eprint {arXiv}
  {1602.07289})

\bibitem[\protect\citeauthoryear{{K{\"o}rding}, {Colbert}  \&
  {Falcke}}{{K{\"o}rding} et~al.}{2005}]{KoerdingColbertFalcke2005}
{K{\"o}rding} E.,  {Colbert} E.,   {Falcke} H.,  2005, \mn@doi [\aap]
  {10.1051/0004-6361:20042452}, \href
  {http://adsabs.harvard.edu/abs/2005A%26A...436..427K} {436, 427}

\bibitem[\protect\citeauthoryear{{K{\"o}rding}, {Jester}  \&
  {Fender}}{{K{\"o}rding} et~al.}{2006}]{KoerdingJesterFender2006}
{K{\"o}rding} E.~G.,  {Jester} S.,   {Fender} R.,  2006, \mn@doi [\mnras]
  {10.1111/j.1365-2966.2006.10954.x}, \href
  {http://adsabs.harvard.edu/abs/2006MNRAS.372.1366K} {372, 1366}

\bibitem[\protect\citeauthoryear{{Kormendy} \& {Ho}}{{Kormendy} \&
  {Ho}}{2013}]{KormendyHo2013}
{Kormendy} J.,  {Ho} L.~C.,  2013, \mn@doi [\araa]
  {10.1146/annurev-astro-082708-101811}, \href
  {http://adsabs.harvard.edu/abs/2013ARA%26A..51..511K} {51, 511}

\bibitem[\protect\citeauthoryear{{LaMassa} et~al.,}{{LaMassa}
  et~al.}{2015}]{LaMassa+2015}
{LaMassa} S.~M.,  et~al., 2015, \mn@doi [\apj] {10.1088/0004-637X/800/2/144},
  \href {http://ukads.nottingham.ac.uk/abs/2015ApJ...800..144L} {800, 144}

\bibitem[\protect\citeauthoryear{{Lemoine} \& {Pelletier}}{{Lemoine} \&
  {Pelletier}}{2003}]{LemoinePelletier2003}
{Lemoine} M.,  {Pelletier} G.,  2003, \mn@doi [\apjl] {10.1086/376353}, \href
  {http://adsabs.harvard.edu/abs/2003ApJ...589L..73L} {589, L73}

\bibitem[\protect\citeauthoryear{{Maitra}, {Markoff}, {Brocksopp}, {Noble},
  {Nowak}  \& {Wilms}}{{Maitra} et~al.}{2009a}]{MaitraMarkoffBrocksopp2009}
{Maitra} D.,  {Markoff} S.,  {Brocksopp} C.,  {Noble} M.,  {Nowak} M.,
  {Wilms} J.,  2009a, \mn@doi [\mnras] {10.1111/j.1365-2966.2009.14896.x},
  \href {http://adsabs.harvard.edu/abs/2009MNRAS.398.1638M} {398, 1638}

\bibitem[\protect\citeauthoryear{{Maitra}, {Markoff}  \& {Falcke}}{{Maitra}
  et~al.}{2009b}]{Maitra+2009b}
{Maitra} D.,  {Markoff} S.,   {Falcke} H.,  2009b, \aap, \href
  {http://cdsads.u-strasbg.fr/abs/2009arXiv0911.0739M} {508}

\bibitem[\protect\citeauthoryear{{Maitra}, {Miller}, {Markoff}  \&
  {King}}{{Maitra} et~al.}{2011}]{Maitra+2011}
{Maitra} D.,  {Miller} J.~M.,  {Markoff} S.,   {King} A.,  2011, \mn@doi [\apj]
  {10.1088/0004-637X/735/2/107}, \href
  {http://adsabs.harvard.edu/abs/2011ApJ...735..107M} {735, 107}

\bibitem[\protect\citeauthoryear{{Makishima}, {Maejima}, {Mitsuda}, {Bradt},
  {Remillard}, {Tuohy}, {Hoshi}  \& {Nakagawa}}{{Makishima}
  et~al.}{1986}]{Makishima1986}
{Makishima} K.,  {Maejima} Y.,  {Mitsuda} K.,  {Bradt} H.~V.,  {Remillard}
  R.~A.,  {Tuohy} I.~R.,  {Hoshi} R.,   {Nakagawa} M.,  1986, \mn@doi [\apj]
  {10.1086/164534}, \href {http://adsabs.harvard.edu/abs/1986ApJ...308..635M}
  {308, 635}

\bibitem[\protect\citeauthoryear{{Malzac}, {Beloborodov}  \&
  {Poutanen}}{{Malzac} et~al.}{2001}]{Malzac+2001}
{Malzac} J.,  {Beloborodov} A.~M.,   {Poutanen} J.,  2001, \mn@doi [\mnras]
  {10.1046/j.1365-8711.2001.04450.x}, \href
  {http://adsabs.harvard.edu/abs/2001MNRAS.326..417M} {326, 417}

\bibitem[\protect\citeauthoryear{{Malzac}, {Belmont}  \& {Fabian}}{{Malzac}
  et~al.}{2009}]{MalzacBelmontFabian2009}
{Malzac} J.,  {Belmont} R.,   {Fabian} A.~C.,  2009, \mn@doi [\mnras]
  {10.1111/j.1365-2966.2009.15553.x}, \href
  {http://cdsads.u-strasbg.fr/abs/2009MNRAS.tmp.1453M} {pp 1453--+}

\bibitem[\protect\citeauthoryear{{Maoz}, {Filippenko}, {Ho}, {Rix}, {Bahcall},
  {Schneider}  \& {Macchetto}}{{Maoz} et~al.}{1995}]{Maoz+1995}
{Maoz} D.,  {Filippenko} A.~V.,  {Ho} L.~C.,  {Rix} H.-W.,  {Bahcall} J.~N.,
  {Schneider} D.~P.,   {Macchetto} F.~D.,  1995, \mn@doi [\apj]
  {10.1086/175250}, \href {http://adsabs.harvard.edu/abs/1995ApJ...440...91M}
  {440, 91}

\bibitem[\protect\citeauthoryear{{Maoz}, {Nagar}, {Falcke}  \& {Wilson}}{{Maoz}
  et~al.}{2005}]{Maoz+2005}
{Maoz} D.,  {Nagar} N.~M.,  {Falcke} H.,   {Wilson} A.~S.,  2005, \mn@doi
  [\apj] {10.1086/429795}, \href
  {http://adsabs.harvard.edu/abs/2005ApJ...625..699M} {625, 699}

\bibitem[\protect\citeauthoryear{{Markoff}, {Falcke}, {Yuan}  \&
  {Biermann}}{{Markoff} et~al.}{2001}]{Markoff+2001}
{Markoff} S.,  {Falcke} H.,  {Yuan} F.,   {Biermann} P.~L.,  2001, \mn@doi
  [\aap] {10.1051/0004-6361:20011346}, \href
  {http://adsabs.harvard.edu/abs/2001A%26A...379L..13M} {379, L13}

\bibitem[\protect\citeauthoryear{{Markoff}, {Nowak}  \& {Wilms}}{{Markoff}
  et~al.}{2005}]{MarkoffNowakWilms2005}
{Markoff} S.,  {Nowak} M.~A.,   {Wilms} J.,  2005, \mn@doi [\apj]
  {10.1086/497628}, \href {http://adsabs.harvard.edu/abs/2005ApJ...635.1203M}
  {635, 1203}

\bibitem[\protect\citeauthoryear{{Markoff}, {Bower}  \& {Falcke}}{{Markoff}
  et~al.}{2007}]{MarkoffBowerFalcke2007}
{Markoff} S.,  {Bower} G.~C.,   {Falcke} H.,  2007, \mn@doi [\mnras]
  {10.1111/j.1365-2966.2007.12071.x}, \href
  {http://adsabs.harvard.edu/abs/2007MNRAS.379.1519M} {379, 1519}

\bibitem[\protect\citeauthoryear{{Markoff} et~al.,}{{Markoff}
  et~al.}{2008}]{Markoff+2008}
{Markoff} S.,  et~al., 2008, \mn@doi [\apj] {10.1086/588718}, \href
  {http://adsabs.harvard.edu/abs/2008ApJ...681..905M} {681, 905}

\bibitem[\protect\citeauthoryear{{Markoff} et~al.,}{{Markoff}
  et~al.}{2015}]{Markoff+2015}
{Markoff} S.,  et~al., 2015, \mn@doi [\apjl] {10.1088/2041-8205/812/2/L25},
  \href {http://adsabs.harvard.edu/abs/2015ApJ...812L..25M} {812, L25}

\bibitem[\protect\citeauthoryear{{Mastichiadis} \& {Ozernoy}}{{Mastichiadis} \&
  {Ozernoy}}{1994}]{MastichiadisOzernoy1994}
{Mastichiadis} A.,  {Ozernoy} L.~M.,  1994, \mn@doi [\apj] {10.1086/174096},
  \href {http://adsabs.harvard.edu/abs/1994ApJ...426..599M} {426, 599}

\bibitem[\protect\citeauthoryear{{McHardy}, {Koerding}, {Knigge}, {Uttley}  \&
  {Fender}}{{McHardy} et~al.}{2006}]{McHardy+2006}
{McHardy} I.~M.,  {Koerding} E.,  {Knigge} C.,  {Uttley} P.,   {Fender} R.~P.,
  2006, \mn@doi [\nat] {10.1038/nature05389}, \href
  {http://adsabs.harvard.edu/abs/2006Natur.444..730M} {444, 730}

\bibitem[\protect\citeauthoryear{{Merloni} \& {Fabian}}{{Merloni} \&
  {Fabian}}{2002}]{MerloniFabian2002}
{Merloni} A.,  {Fabian} A.~C.,  2002, \mn@doi [\mnras]
  {10.1046/j.1365-8711.2002.05288.x}, \href
  {http://adsabs.harvard.edu/abs/2002MNRAS.332..165M} {332, 165}

\bibitem[\protect\citeauthoryear{{Merloni}, {Heinz}  \& {di Matteo}}{{Merloni}
  et~al.}{2003}]{MerloniHeinzDiMatteo2003}
{Merloni} A.,  {Heinz} S.,   {di Matteo} T.,  2003, \mnras, \href
  {http://adsabs.harvard.edu/cgi-bin/nph-bib_query?bibcode=2003MNRAS.345.1057M&amp;db_key=AST}
  {345, 1057}

\bibitem[\protect\citeauthoryear{{Metzger}, {Williams}  \& {Berger}}{{Metzger}
  et~al.}{2015}]{Metzger+2015}
{Metzger} B.~D.,  {Williams} P.~K.~G.,   {Berger} E.,  2015, \mn@doi [\apj]
  {10.1088/0004-637X/806/2/224}, \href
  {http://adsabs.harvard.edu/abs/2015ApJ...806..224M} {806, 224}

\bibitem[\protect\citeauthoryear{{Migliari} et~al.,}{{Migliari}
  et~al.}{2007}]{Migliari+2007}
{Migliari} S.,  et~al., 2007, \mn@doi [\apj] {10.1086/522023}, \href
  {http://adsabs.harvard.edu/abs/2007ApJ...670..610M} {670, 610}

\bibitem[\protect\citeauthoryear{{Mitsuda} et~al.,}{{Mitsuda}
  et~al.}{1984}]{Mitsuda+1984}
{Mitsuda} K.,  et~al., 1984, \pasj, \href
  {http://adsabs.harvard.edu/abs/1984PASJ...36..741M} {36, 741}

\bibitem[\protect\citeauthoryear{{Miyakawa}, {Yamaoka}, {Homan}, {Saito},
  {Dotani}, {Yoshida}  \& {Inoue}}{{Miyakawa} et~al.}{2008}]{Miyakawa+2008}
{Miyakawa} T.,  {Yamaoka} K.,  {Homan} J.,  {Saito} K.,  {Dotani} T.,
  {Yoshida} A.,   {Inoue} H.,  2008, \mn@doi [\pasj] {10.1093/pasj/60.3.637},
  \href {http://adsabs.harvard.edu/abs/2008PASJ...60..637M} {60, 637}

\bibitem[\protect\citeauthoryear{{Mo{\'s}cibrodzka}, {Falcke}  \&
  {Shiokawa}}{{Mo{\'s}cibrodzka} et~al.}{2016}]{Moscibrodzka+2016}
{Mo{\'s}cibrodzka} M.,  {Falcke} H.,   {Shiokawa} H.,  2016, \mn@doi [\aap]
  {10.1051/0004-6361/201526630}, \href
  {http://adsabs.harvard.edu/abs/2016A%26A...586A..38M} {586, A38}

\bibitem[\protect\citeauthoryear{{Moustakas} \& {Kennicutt}}{{Moustakas} \&
  {Kennicutt}}{2006}]{MoustakasKennicutt2006}
{Moustakas} J.,  {Kennicutt} Jr. R.~C.,  2006, \mn@doi [\apjs]
  {10.1086/500971}, \href {http://adsabs.harvard.edu/abs/2006ApJS..164...81M}
  {164, 81}

\bibitem[\protect\citeauthoryear{{Mullaney}, {Alexander}, {Fine}, {Goulding},
  {Harrison}  \& {Hickox}}{{Mullaney} et~al.}{2013}]{Mullaney+2013}
{Mullaney} J.~R.,  {Alexander} D.~M.,  {Fine} S.,  {Goulding} A.~D.,
  {Harrison} C.~M.,   {Hickox} R.~C.,  2013, \mn@doi [\mnras]
  {10.1093/mnras/stt751}, \href
  {http://adsabs.harvard.edu/abs/2013MNRAS.433..622M} {433, 622}

\bibitem[\protect\citeauthoryear{{Nagar}, {Falcke}, {Wilson}  \&
  {Ulvestad}}{{Nagar} et~al.}{2002}]{NagarFalckeWilson2002}
{Nagar} N.~M.,  {Falcke} H.,  {Wilson} A.~S.,   {Ulvestad} J.~S.,  2002,
  \mn@doi [\aap] {10.1051/0004-6361:20020874}, \href
  {http://adsabs.harvard.edu/abs/2002A%26A...392...53N} {392, 53}

\bibitem[\protect\citeauthoryear{{Nagar}, {Falcke}  \& {Wilson}}{{Nagar}
  et~al.}{2005}]{NagarFalckeWilson2005}
{Nagar} N.~M.,  {Falcke} H.,   {Wilson} A.~S.,  2005, \mn@doi [\aap]
  {10.1051/0004-6361:20042277}, \href
  {http://adsabs.harvard.edu/abs/2005A%26A...435..521N} {435, 521}

\bibitem[\protect\citeauthoryear{{Neilsen} \& {Lee}}{{Neilsen} \&
  {Lee}}{2009}]{NeilsenLee2009}
{Neilsen} J.,  {Lee} J.~C.,  2009, \mn@doi [\nat] {10.1038/nature07680}, \href
  {http://adsabs.harvard.edu/abs/2009Natur.458..481N} {458, 481}

\bibitem[\protect\citeauthoryear{{Neilsen} et~al.,}{{Neilsen}
  et~al.}{2013}]{Neilsen+2013}
{Neilsen} J.,  et~al., 2013, \mn@doi [\apj] {10.1088/0004-637X/774/1/42}, \href
  {http://adsabs.harvard.edu/abs/2013ApJ...774...42N} {774, 42}

\bibitem[\protect\citeauthoryear{{Nemmen} \& {Tchekhovskoy}}{{Nemmen} \&
  {Tchekhovskoy}}{2015}]{NemmenTchekhovskoy2015}
{Nemmen} R.~S.,  {Tchekhovskoy} A.,  2015, \mn@doi [\mnras]
  {10.1093/mnras/stv260}, \href
  {http://adsabs.harvard.edu/abs/2015MNRAS.449..316N} {449, 316}

\bibitem[\protect\citeauthoryear{{Nemmen}, {Storchi-Bergmann}  \&
  {Eracleous}}{{Nemmen} et~al.}{2014}]{Nemmen+2014}
{Nemmen} R.~S.,  {Storchi-Bergmann} T.,   {Eracleous} M.,  2014, \mn@doi
  [\mnras] {10.1093/mnras/stt2388}, \href
  {http://adsabs.harvard.edu/abs/2014MNRAS.438.2804N} {438, 2804}

\bibitem[\protect\citeauthoryear{{Nesvadba}, {Lehnert}, {De Breuck}, {Gilbert}
  \& {van Breugel}}{{Nesvadba} et~al.}{2008}]{Nesvadba+2008}
{Nesvadba} N.~P.~H.,  {Lehnert} M.~D.,  {De Breuck} C.,  {Gilbert} A.~M.,
  {van Breugel} W.,  2008, \mn@doi [\aap] {10.1051/0004-6361:200810346}, \href
  {http://adsabs.harvard.edu/abs/2008A%26A...491..407N} {491, 407}

\bibitem[\protect\citeauthoryear{{Netzer}}{{Netzer}}{2013}]{Netzerbook}
{Netzer} H.,  2013, {The Physics and Evolution of Active Galactic Nuclei}

\bibitem[\protect\citeauthoryear{{Netzer}}{{Netzer}}{2015}]{Netzer2015}
{Netzer} H.,  2015, preprint, \href
  {http://adsabs.harvard.edu/abs/2015arXiv150500811N} {} (\mn@eprint {arXiv}
  {1505.00811})

\bibitem[\protect\citeauthoryear{{Nied{\'z}wiecki}, {Xie}  \& {St{\c
  e}pnik}}{{Nied{\'z}wiecki} et~al.}{2014}]{Niedzwiecki+2014}
{Nied{\'z}wiecki} A.,  {Xie} F.-G.,   {St{\c e}pnik} A.,  2014, \mn@doi
  [\mnras] {10.1093/mnras/stu1262}, \href
  {http://adsabs.harvard.edu/abs/2014MNRAS.443.1733N} {443, 1733}

\bibitem[\protect\citeauthoryear{{Novikov} \& {Thorne}}{{Novikov} \&
  {Thorne}}{1973}]{NovikovThorne1973}
{Novikov} I.~D.,  {Thorne} K.~S.,  1973, in {Dewitt} C.,  {Dewitt} B.~S.,  eds,
  Black Holes (Les Astres Occlus). pp 343--450

\bibitem[\protect\citeauthoryear{{Pellegrini}, {Fabbiano}, {Fiore},
  {Trinchieri}  \& {Antonelli}}{{Pellegrini} et~al.}{2002}]{Pellegrini+2002}
{Pellegrini} S.,  {Fabbiano} G.,  {Fiore} F.,  {Trinchieri} G.,   {Antonelli}
  A.,  2002, \mn@doi [\aap] {10.1051/0004-6361:20011482}, \href
  {http://adsabs.harvard.edu/abs/2002A%26A...383....1P} {383, 1}

\bibitem[\protect\citeauthoryear{{Pian}, {Romano}, {Maoz}, {Cucchiara},
  {Pagani}  \& {Parola}}{{Pian} et~al.}{2010}]{Pian+2010}
{Pian} E.,  {Romano} P.,  {Maoz} D.,  {Cucchiara} A.,  {Pagani} C.,   {Parola}
  V.~L.,  2010, \mn@doi [\mnras] {10.1111/j.1365-2966.2009.15689.x}, \href
  {http://adsabs.harvard.edu/abs/2010MNRAS.401..677P} {401, 677}

\bibitem[\protect\citeauthoryear{{Plotkin}, {Markoff}, {Kelly}, {K{\"o}rding}
  \& {Anderson}}{{Plotkin} et~al.}{2012}]{Plotkin+2012}
{Plotkin} R.~M.,  {Markoff} S.,  {Kelly} B.~C.,  {K{\"o}rding} E.,   {Anderson}
  S.~F.,  2012, \mn@doi [\mnras] {10.1111/j.1365-2966.2011.19689.x}, \href
  {http://adsabs.harvard.edu/abs/2012MNRAS.419..267P} {419, 267}

\bibitem[\protect\citeauthoryear{{Plotkin}, {Gallo}, {Markoff}, {Homan},
  {Jonker}, {Miller-Jones}, {Russell}  \& {Drappeau}}{{Plotkin}
  et~al.}{2015}]{Plotkin+2015}
{Plotkin} R.~M.,  {Gallo} E.,  {Markoff} S.,  {Homan} J.,  {Jonker} P.~G.,
  {Miller-Jones} J.~C.~A.,  {Russell} D.~M.,   {Drappeau} S.,  2015, \mn@doi
  [\mnras] {10.1093/mnras/stu2385}, \href
  {http://adsabs.harvard.edu/abs/2015MNRAS.446.4098P} {446, 4098}

\bibitem[\protect\citeauthoryear{{Polko}, {Meier}  \& {Markoff}}{{Polko}
  et~al.}{2010}]{Polko+2010}
{Polko} P.,  {Meier} D.~L.,   {Markoff} S.,  2010, \mn@doi [\apj]
  {10.1088/0004-637X/723/2/1343}, \href
  {http://adsabs.harvard.edu/abs/2010ApJ...723.1343P} {723, 1343}

\bibitem[\protect\citeauthoryear{{Polko}, {Meier}  \& {Markoff}}{{Polko}
  et~al.}{2013}]{Polko+2013}
{Polko} P.,  {Meier} D.~L.,   {Markoff} S.,  2013, \mn@doi [\mnras]
  {10.1093/mnras/sts052}, \href
  {http://adsabs.harvard.edu/abs/2013MNRAS.428..587P} {428, 587}

\bibitem[\protect\citeauthoryear{{Polko}, {Meier}  \& {Markoff}}{{Polko}
  et~al.}{2014}]{Polko+2014}
{Polko} P.,  {Meier} D.~L.,   {Markoff} S.,  2014, \mn@doi [\mnras]
  {10.1093/mnras/stt2155}, \href
  {http://adsabs.harvard.edu/abs/2014MNRAS.438..959P} {438, 959}

\bibitem[\protect\citeauthoryear{{Ponti}, {Fender}, {Begelman}, {Dunn},
  {Neilsen}  \& {Coriat}}{{Ponti} et~al.}{2012}]{Ponti+2012}
{Ponti} G.,  {Fender} R.~P.,  {Begelman} M.~C.,  {Dunn} R.~J.~H.,  {Neilsen}
  J.,   {Coriat} M.,  2012, \mn@doi [\mnras]
  {10.1111/j.1745-3933.2012.01224.x}, \href
  {http://adsabs.harvard.edu/abs/2012MNRAS.422L..11P} {422, 11}

\bibitem[\protect\citeauthoryear{{Prieto}, {Fern{\'a}ndez-Ontiveros},
  {Markoff}, {Espada}  \& {Gonz{\'a}lez-Mart{\'{\i}}n}}{{Prieto}
  et~al.}{2016}]{Prieto+2016}
{Prieto} M.~A.,  {Fern{\'a}ndez-Ontiveros} J.~A.,  {Markoff} S.,  {Espada} D.,
   {Gonz{\'a}lez-Mart{\'{\i}}n} O.,  2016, \mn@doi [\mnras]
  {10.1093/mnras/stw166}, \href
  {http://adsabs.harvard.edu/abs/2016MNRAS.457.3801P} {457, 3801}

\bibitem[\protect\citeauthoryear{{Remillard} \& {McClintock}}{{Remillard} \&
  {McClintock}}{2006}]{RemillardMcClintock2006}
{Remillard} R.~A.,  {McClintock} J.~E.,  2006, \mn@doi [\araa]
  {10.1146/annurev.astro.44.051905.092532PDF:
  http://arjournals.annualreviews.org/doi/pdf/10.1146/annurev.astro.44.051905.092532},
  \href {http://adsabs.harvard.edu/abs/2006ARA%26A..44...49R} {44, 49}

\bibitem[\protect\citeauthoryear{{Richings}, {Uttley}  \&
  {K{\"o}rding}}{{Richings} et~al.}{2011}]{RichingsUttleyKording2011}
{Richings} A.~J.,  {Uttley} P.,   {K{\"o}rding} E.,  2011, \mn@doi [\mnras]
  {10.1111/j.1365-2966.2011.18845.x}, \href
  {http://adsabs.harvard.edu/abs/2011MNRAS.415.2158R} {415, 2158}

\bibitem[\protect\citeauthoryear{{Roberts}, {Warwick}  \& {Ohashi}}{{Roberts}
  et~al.}{1999}]{Roberts+1999}
{Roberts} T.~P.,  {Warwick} R.~S.,   {Ohashi} T.,  1999, \mn@doi [\mnras]
  {10.1046/j.1365-8711.1999.02359.x}, \href
  {http://adsabs.harvard.edu/abs/1999MNRAS.304...52R} {304, 52}

\bibitem[\protect\citeauthoryear{{Roberts}, {Schurch}  \& {Warwick}}{{Roberts}
  et~al.}{2001}]{Roberts+2001}
{Roberts} T.~P.,  {Schurch} N.~J.,   {Warwick} R.~S.,  2001, \mn@doi [\mnras]
  {10.1046/j.1365-8711.2001.04365.x}, \href
  {http://adsabs.harvard.edu/abs/2001MNRAS.324..737R} {324, 737}

\bibitem[\protect\citeauthoryear{{Russell}, {Maitra}, {Dunn}  \&
  {Markoff}}{{Russell} et~al.}{2010}]{Russell+2010}
{Russell} D.~M.,  {Maitra} D.,  {Dunn} R.~J.~H.,   {Markoff} S.,  2010, \mn@doi
  [\mnras] {10.1111/j.1365-2966.2010.16547.x}, \href
  {http://adsabs.harvard.edu/abs/2010MNRAS.405.1759R} {405, 1759}

\bibitem[\protect\citeauthoryear{{Russell} et~al.,}{{Russell}
  et~al.}{2013}]{Russell+2013}
{Russell} D.~M.,  et~al., 2013, \mn@doi [\mnras] {10.1093/mnras/sts377}, \href
  {http://adsabs.harvard.edu/abs/2013MNRAS.429..815R} {429, 815}

\bibitem[\protect\citeauthoryear{{Schawinski}, {Thomas}, {Sarzi}, {Maraston},
  {Kaviraj}, {Joo}, {Yi}  \& {Silk}}{{Schawinski}
  et~al.}{2007}]{Schawinski+2007}
{Schawinski} K.,  {Thomas} D.,  {Sarzi} M.,  {Maraston} C.,  {Kaviraj} S.,
  {Joo} S.-J.,  {Yi} S.~K.,   {Silk} J.,  2007, \mn@doi [\mnras]
  {10.1111/j.1365-2966.2007.12487.x}, \href
  {http://adsabs.harvard.edu/abs/2007MNRAS.382.1415S} {382, 1415}

\bibitem[\protect\citeauthoryear{{Schawinski} et~al.,}{{Schawinski}
  et~al.}{2010}]{Schawinski+2010}
{Schawinski} K.,  et~al., 2010, \mn@doi [\apjl] {10.1088/2041-8205/724/1/L30},
  \href {http://adsabs.harvard.edu/abs/2010ApJ...724L..30S} {724, L30}

\bibitem[\protect\citeauthoryear{{Schawinski}, {Koss}, {Berney}  \&
  {Sartori}}{{Schawinski} et~al.}{2015}]{Schawinski+2015}
{Schawinski} K.,  {Koss} M.,  {Berney} S.,   {Sartori} L.~F.,  2015, \mn@doi
  [\mnras] {10.1093/mnras/stv1136}, \href
  {http://adsabs.harvard.edu/abs/2015MNRAS.451.2517S} {451, 2517}

\bibitem[\protect\citeauthoryear{{Seaton}}{{Seaton}}{1979}]{Seaton1979}
{Seaton} M.~J.,  1979, \mnras, \href
  {http://adsabs.harvard.edu/abs/1979MNRAS.187P..73S} {187, 73P}

\bibitem[\protect\citeauthoryear{{Seyfert}}{{Seyfert}}{1943}]{Seyfert1943}
{Seyfert} C.~K.,  1943, \mn@doi [\apj] {10.1086/144488}, \href
  {http://adsabs.harvard.edu/abs/1943ApJ....97...28S} {97, 28}

\bibitem[\protect\citeauthoryear{{Shakura} \& {Sunyaev}}{{Shakura} \&
  {Sunyaev}}{1973}]{ShakuraSunyaev1973}
{Shakura} N.~I.,  {Sunyaev} R.~A.,  1973, \aap, \href
  {http://adsabs.harvard.edu/abs/1973A%26A....24..337S} {24, 337}

\bibitem[\protect\citeauthoryear{{Sikora} \& {Begelman}}{{Sikora} \&
  {Begelman}}{2013}]{SikoraBegelman2013}
{Sikora} M.,  {Begelman} M.~C.,  2013, \mn@doi [\apjl]
  {10.1088/2041-8205/764/2/L24}, \href
  {http://adsabs.harvard.edu/abs/2013ApJ...764L..24S} {764, L24}

\bibitem[\protect\citeauthoryear{{Stark}, {Gammie}, {Wilson}, {Bally}, {Linke},
  {Heiles}  \& {Hurwitz}}{{Stark} et~al.}{1992}]{Stark+1992}
{Stark} A.~A.,  {Gammie} C.~F.,  {Wilson} R.~W.,  {Bally} J.,  {Linke} R.~A.,
  {Heiles} C.,   {Hurwitz} M.,  1992, \mn@doi [\apjs] {10.1086/191645}, \href
  {http://adsabs.harvard.edu/abs/1992ApJS...79...77S} {79, 77}

\bibitem[\protect\citeauthoryear{{Sunyaev} \& {Truemper}}{{Sunyaev} \&
  {Truemper}}{1979}]{SunyaevTruemper1979}
{Sunyaev} R.~A.,  {Truemper} J.,  1979, \mn@doi [\nat] {10.1038/279506a0},
  \href {http://adsabs.harvard.edu/abs/1979Natur.279..506S} {279, 506}

\bibitem[\protect\citeauthoryear{{Tadhunter}}{{Tadhunter}}{2016}]{Tadhunter2016}
{Tadhunter} C.,  2016, \mn@doi [\aapr] {10.1007/s00159-016-0094-x}, \href
  {http://adsabs.harvard.edu/abs/2016A%26ARv..24...10T} {24, 10}

\bibitem[\protect\citeauthoryear{{Tchekhovskoy}, {Narayan}  \&
  {McKinney}}{{Tchekhovskoy} et~al.}{2011}]{Tchekhovskoy+2011}
{Tchekhovskoy} A.,  {Narayan} R.,   {McKinney} J.~C.,  2011, \mn@doi [\mnras]
  {10.1111/j.1745-3933.2011.01147.x}, \href
  {http://adsabs.harvard.edu/abs/2011MNRAS.418L..79T} {418, L79}

\bibitem[\protect\citeauthoryear{{Thorne} \& {Price}}{{Thorne} \&
  {Price}}{1975}]{ThornePrice1975}
{Thorne} K.~S.,  {Price} R.~H.,  1975, \mn@doi [\apjl] {10.1086/181720}, \href
  {http://adsabs.harvard.edu/abs/1975ApJ...195L.101T} {195, L101}

\bibitem[\protect\citeauthoryear{{Tonry}, {Dressler}, {Blakeslee}, {Ajhar},
  {Fletcher}, {Luppino}, {Metzger}  \& {Moore}}{{Tonry}
  et~al.}{2001}]{Tonry+2001}
{Tonry} J.~L.,  {Dressler} A.,  {Blakeslee} J.~P.,  {Ajhar} E.~A.,  {Fletcher}
  A.~B.,  {Luppino} G.~A.,  {Metzger} M.~R.,   {Moore} C.~B.,  2001, \mn@doi
  [\apj] {10.1086/318301}, \href
  {http://adsabs.harvard.edu/abs/2001ApJ...546..681T} {546, 681}

\bibitem[\protect\citeauthoryear{{Trujillo}, {Martinez-Valpuesta},
  {Mart{\'{\i}}nez-Delgado}, {Pe{\~n}arrubia}, {Gabany}  \&
  {Pohlen}}{{Trujillo} et~al.}{2009}]{Trujillo+2009}
{Trujillo} I.,  {Martinez-Valpuesta} I.,  {Mart{\'{\i}}nez-Delgado} D.,
  {Pe{\~n}arrubia} J.,  {Gabany} R.~J.,   {Pohlen} M.,  2009, \mn@doi [\apj]
  {10.1088/0004-637X/704/1/618}, \href
  {http://adsabs.harvard.edu/abs/2009ApJ...704..618T} {704, 618}

\bibitem[\protect\citeauthoryear{{Ulvestad} \& {Wilson}}{{Ulvestad} \&
  {Wilson}}{1989}]{UlvestadWilson1989}
{Ulvestad} J.~S.,  {Wilson} A.~S.,  1989, \mn@doi [\apj] {10.1086/167737},
  \href {http://adsabs.harvard.edu/abs/1989ApJ...343..659U} {343, 659}

\bibitem[\protect\citeauthoryear{{Veilleux} \& {Osterbrock}}{{Veilleux} \&
  {Osterbrock}}{1987}]{VeilleuxOsterbrock1987}
{Veilleux} S.,  {Osterbrock} D.~E.,  1987, \mn@doi [\apjs] {10.1086/191166},
  \href {http://adsabs.harvard.edu/abs/1987ApJS...63..295V} {63, 295}

\bibitem[\protect\citeauthoryear{{Xie} \& {Yuan}}{{Xie} \&
  {Yuan}}{2016}]{XieYuan2016}
{Xie} F.-G.,  {Yuan} F.,  2016, \mn@doi [\mnras] {10.1093/mnras/stv2956}, \href
  {http://adsabs.harvard.edu/abs/2016MNRAS.456.4377X} {456, 4377}

\bibitem[\protect\citeauthoryear{{Younes}, {Porquet}, {Sabra}  \&
  {Reeves}}{{Younes} et~al.}{2011}]{Younes+2011}
{Younes} G.,  {Porquet} D.,  {Sabra} B.,   {Reeves} J.~N.,  2011, \mn@doi
  [\aap] {10.1051/0004-6361/201116806}, \href
  {http://adsabs.harvard.edu/abs/2011A%26A...530A.149Y} {530, A149}

\bibitem[\protect\citeauthoryear{{Yuan}, {Markoff}  \& {Falcke}}{{Yuan}
  et~al.}{2002a}]{Yuan+2002}
{Yuan} F.,  {Markoff} S.,   {Falcke} H.,  2002a, \mn@doi [\aap]
  {10.1051/0004-6361:20011709}, \href
  {http://adsabs.harvard.edu/abs/2002A%26A...383..854Y} {383, 854}

\bibitem[\protect\citeauthoryear{{Yuan}, {Markoff}, {Falcke}  \&
  {Biermann}}{{Yuan} et~al.}{2002b}]{Yuan+2002_4258}
{Yuan} F.,  {Markoff} S.,  {Falcke} H.,   {Biermann} P.~L.,  2002b, \mn@doi
  [\aap] {10.1051/0004-6361:20020817}, \href
  {http://adsabs.harvard.edu/abs/2002A%26A...391..139Y} {391, 139}

\bibitem[\protect\citeauthoryear{{Yuan}, {Taam}, {Xue}  \& {Cui}}{{Yuan}
  et~al.}{2006}]{Yuan+2006}
{Yuan} F.,  {Taam} R.~E.,  {Xue} Y.,   {Cui} W.,  2006, \mn@doi [\apj]
  {10.1086/497980}, \href {http://adsabs.harvard.edu/abs/2006ApJ...636...46Y}
  {636, 46}

\bibitem[\protect\citeauthoryear{{de Blok}, {Walter}, {Brinks}, {Trachternach},
  {Oh}  \& {Kennicutt}}{{de Blok} et~al.}{2008}]{deBlok+2008}
{de Blok} W.~J.~G.,  {Walter} F.,  {Brinks} E.,  {Trachternach} C.,  {Oh}
  S.-H.,   {Kennicutt} Jr. R.~C.,  2008, \mn@doi [\aj]
  {10.1088/0004-6256/136/6/2648}, \href
  {http://adsabs.harvard.edu/abs/2008AJ....136.2648D} {136, 2648}

\bibitem[\protect\citeauthoryear{{van Oers} et~al.,}{{van Oers}
  et~al.}{2010}]{vanOers+2010}
{van Oers} P.,  et~al., 2010, \mn@doi [\mnras]
  {10.1111/j.1365-2966.2010.17339.x}, \href
  {http://adsabs.harvard.edu/abs/2010MNRAS.409..763V} {409, 763}

\bibitem[\protect\citeauthoryear{{van der Laan} et~al.,}{{van der Laan}
  et~al.}{2015}]{vanderLaan+2015}
{van der Laan} T.~P.~R.,  et~al., 2015, \mn@doi [\aap]
  {10.1051/0004-6361/201425402}, \href
  {http://adsabs.harvard.edu/abs/2015A%26A...575A..83V} {575, A83}

\makeatother
\end{thebibliography}

\appendix
\section{\emerlin contour plots}
\renewcommand\thefigure{\thesection.\arabic{figure}} 
\setcounter{figure}{0}   

\begin{figure*}
\centering
\includegraphics[width=.495\textwidth]{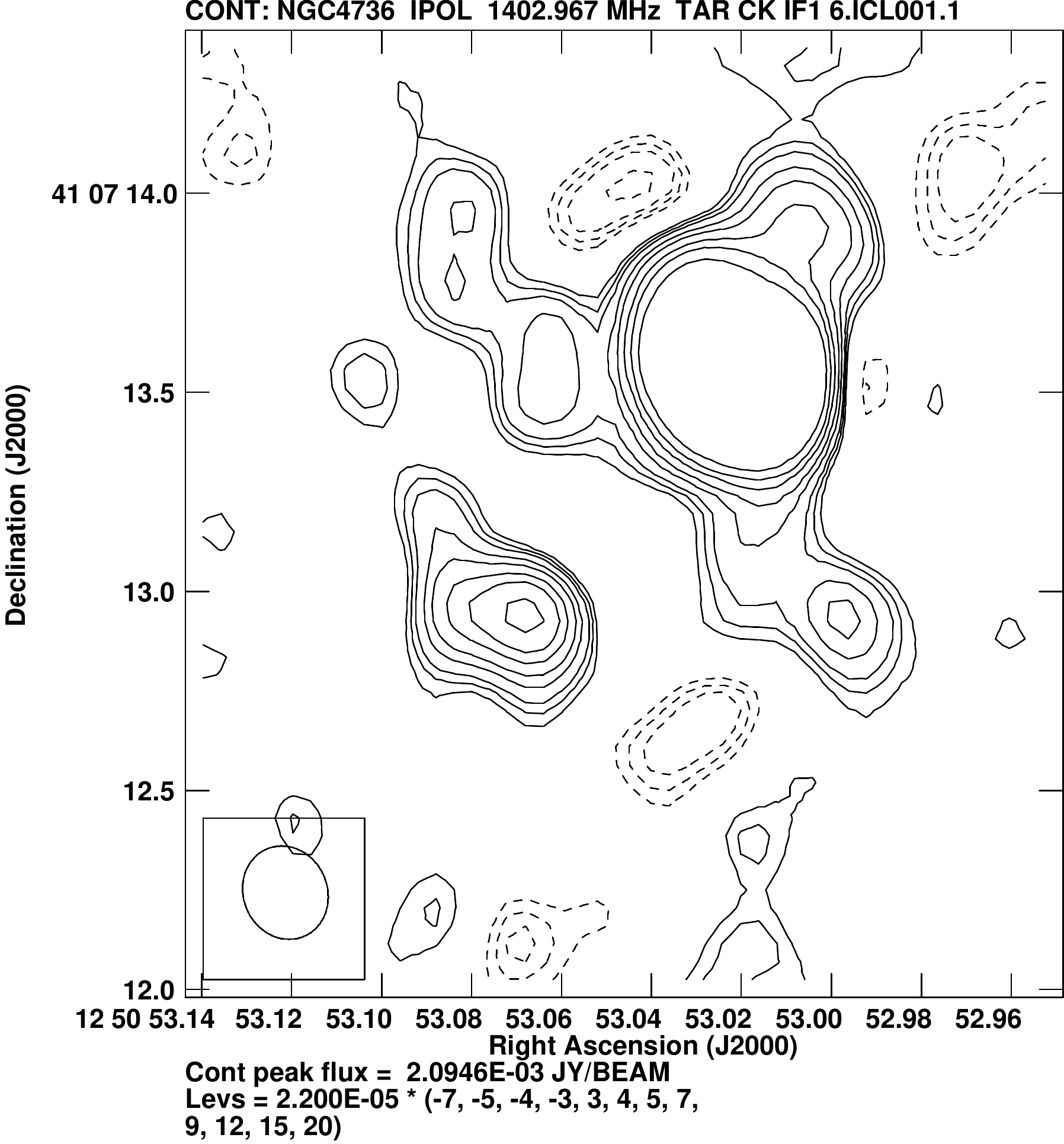}
\includegraphics[width=.495\textwidth]{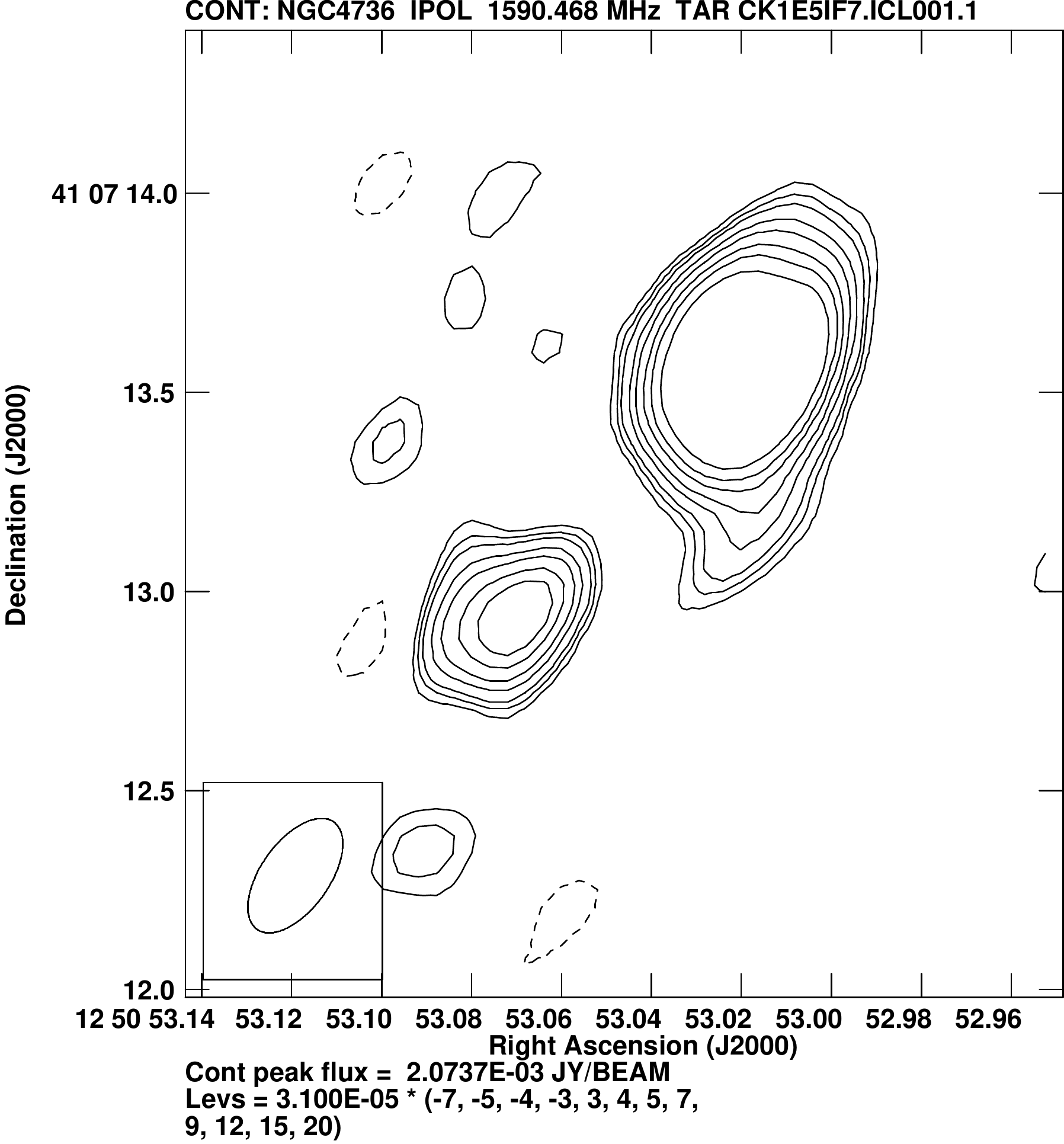}
\includegraphics[width=.495\textwidth]{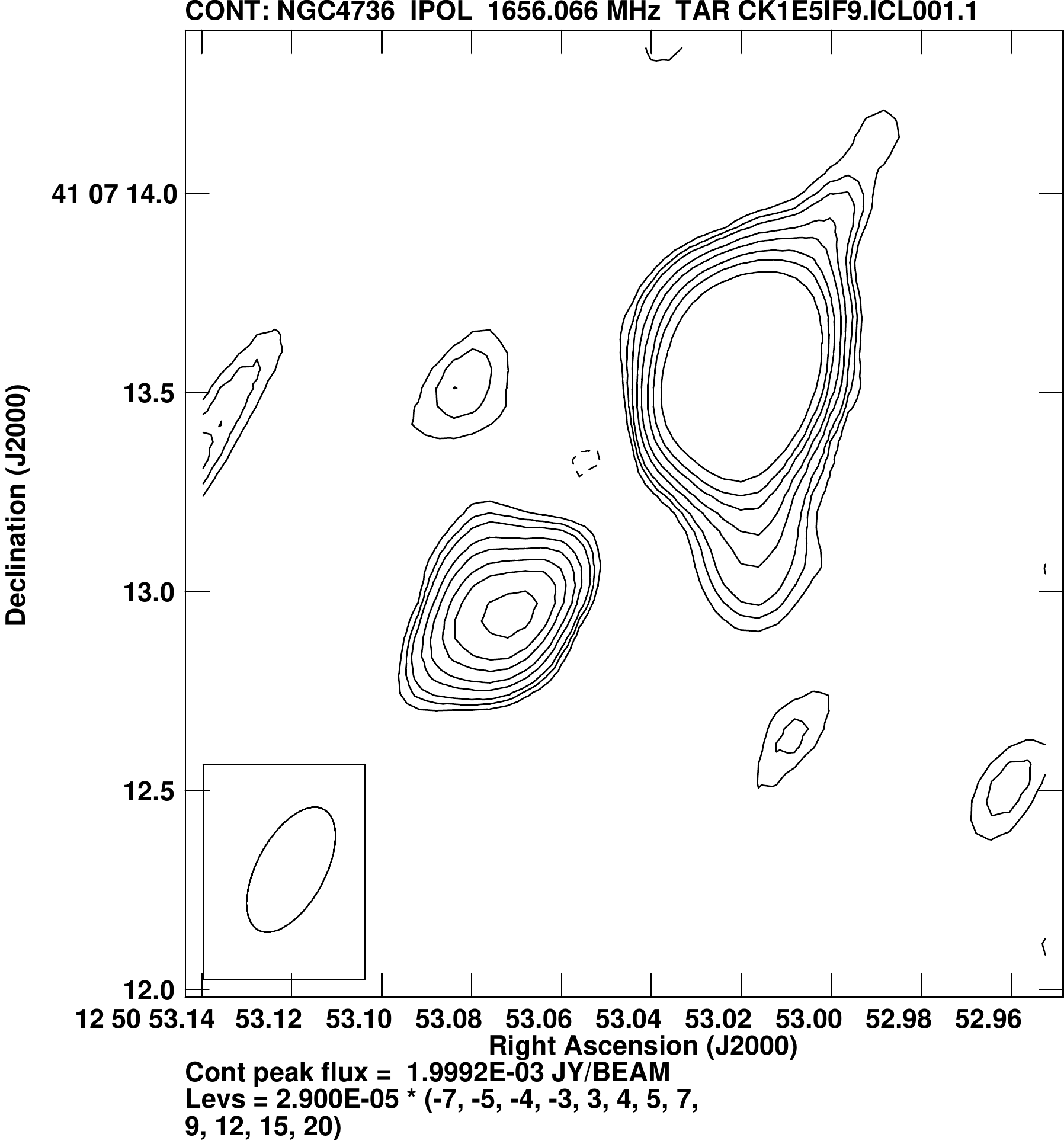}
\includegraphics[width=.495\textwidth]{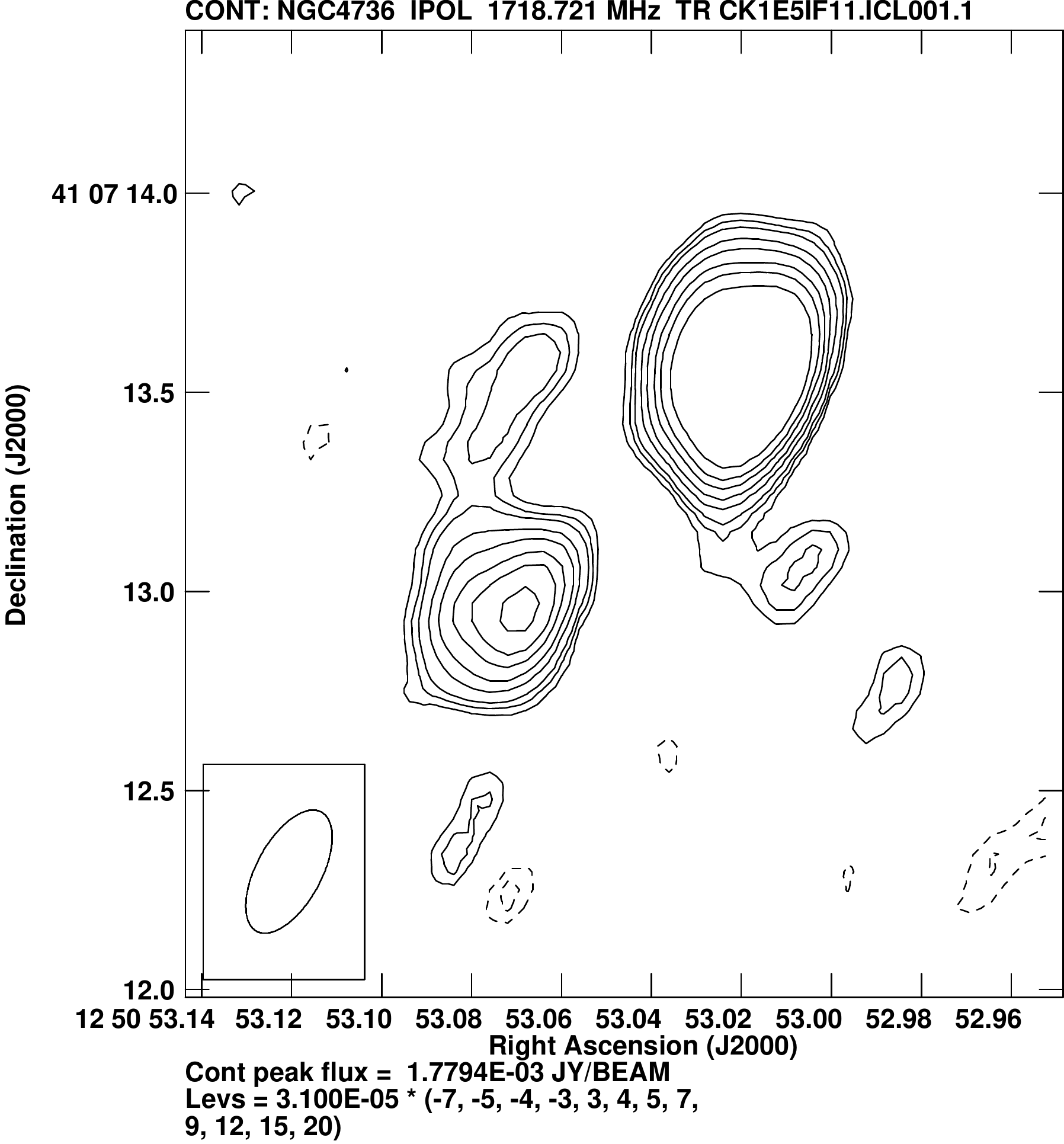}
 \caption{Contour plots of SPWs 1 through 6 binned together (top left) and of SPW 7 (top right), 9 (bottom left) and 11 (bottom right) individually. Contour levels are plotted as specified in the image.}
 \label{fig:kntr}
\end{figure*}

\end{document}